\documentclass[aps,twocolumn,showpacs,preprintnumbers,nofootinbib,superscriptaddress,pdflatex]{revtex4-1}

\usepackage{amsmath,amssymb,bm}
\usepackage[dvips]{graphicx}
\usepackage{graphicx}
\usepackage{hyperref}
\usepackage{yfonts}
\usepackage{color}
\newcommand{\beq}{\begin{eqnarray}}
\newcommand{\eeq}{\end{eqnarray}}

\begin{document}

\title{Chiral magnetohydrodynamic turbulence in core-collapse supernovae}

\author{Youhei Masada}
\affiliation{Department of Science Education, Aichi University of Education, Kariya 448-8542, Japan} 

\author{Kei Kotake}
\affiliation{Faculty of Science, Department of Applied Physics, Fukuoka University, Fukuoka 814-0180, Japan}

\author{Tomoya Takiwaki}
\affiliation{Division of Theoretical Astronomy, National Astronomical Observatory of Japan, Tokyo 181-8588, Japan}

\author{Naoki Yamamoto}
\affiliation{Department of Physics, Keio University, Yokohama 223-8522, Japan}

\begin{abstract}
Macroscopic evolution of relativistic charged matter with chirality imbalance is described by the 
chiral magnetohydrodynamics (chiral MHD).  
One such astrophysical system is high-density lepton matter in core-collapse supernovae 
where the chirality imbalance of leptons is generated by the parity-violating weak processes.
After developing the chiral MHD equations for this system, we perform numerical simulations for the real-time 
evolutions of magnetic and flow fields, and study the properties of the chiral MHD turbulence. 
In particular, we observe the inverse cascade of the magnetic energy and the fluid kinetic energy. 
Our results suggest that the chiral effects that have been neglected so far can reverse the turbulent 
cascade direction from direct to inverse cascade, 
which would impact the magnetohydrodynamics evolution in the supernova core toward explosion.
\end{abstract}
\maketitle

\section{Introduction}
\label{sec:introduction}
Relativistic matter with chirality imbalance (chiral matter, in short) is relevant to various physical systems
from hot electroweak plasmas in the early Universe \cite{Joyce:1997uy,Boyarsky:2011uy}, 
quark-gluon plasmas created in heavy ion collision experiments \cite{Kharzeev:2015znc}, 
dense electromagnetic plasmas in neutron stars
\cite{Charbonneau:2009ax,Akamatsu:2013pjd,Ohnishi:2014uea,Kaminski:2014jda} 
and neutrino matter in core-collapse supernovae \cite{Yamamoto:2015gzz} to emergent chiral matter 
near band crossing points of Weyl (semi)metals \cite{Nielsen:1983rb,Vishwanath,BurkovBalents,Xu-chern}. 
In such chiral matter, anomalous transport phenomena that are absent in usual parity-invariant matter 
emerge. Two prominent examples are the current along the direction of a magnetic field, called the 
chiral magnetic effect (CME) \cite{Vilenkin:1980fu,Nielsen:1983rb,Alekseev:1998ds,Fukushima:2008xe}, 
and the current along the direction of a vorticity, called the chiral vortical effect (CVE) 
\cite{Vilenkin:1979ui,Kharzeev:2007tn,Son:2009tf,Landsteiner:2011cp}. 
Notably, the CME and CVE have close connection with the quantum violation of the chiral symmetry
(or nonconservation of the chiral charge) in relativistic quantum field theories, known as the chiral
anomaly \cite{Adler,BellJackiw}. 
The macroscopic evolution of charged chiral matter at a long time distance and long timescale is then 
described by the hydrodynamic theory by incorporating the effects of the chiral transport phenomena and the
chiral anomaly. 
This is the {\it chiral magnetohydrodynamics} (chiral MHD) \cite{Giovannini:2013oga,Boyarsky:2015faa,
Yamamoto:2015gzz,Yamamoto:2016xtu,Rogachevskii:2017uyc,Hattori:2017usa}.

It is natural to expect that real-time evolution of chiral matter described by the chiral MHD is 
qualitatively different from that of nonchiral matter described by the conventional MHD. 
Analytically tractable regimes of the chiral MHD have been studied in 
Refs.~\cite{Giovannini:2013oga,Boyarsky:2015faa,Yamamoto:2015gzz,Yamamoto:2016xtu,Pavlovic:2016gac,Rogachevskii:2017uyc,Hattori:2017usa}.
Among others, inverse cascade of the fluid kinetic energy, i.e., the energy transfer from large to small scales, 
in addition to the inverse cascade of the magnetic energy \cite{Hirono:2015rla,Gorbar:2016klv},
in the chiral MHD turbulence was predicted for pure chiral matter under certain conditions \cite{Yamamoto:2016xtu}. 
This should be contrasted with the direct energy cascade (the energy transfer from small to large scales) in usual
nonchiral matter.
More recently, chiral MHD equations were numerically studied for high-temperature electroweak plasmas in the early
Universe and inverse energy cascade was indeed observed 
\cite{Schober:2017cdw,Brandenburg:2017rcb,Schober:2018ojn}.%
\footnote{The anomalous hydrodynamics with the CME in {\it external} electromagnetic fields has 
been studied in heavy ion physics~\cite{Hongo:2013cqa,Hirono:2014oda,Jiang:2016wve}.}
A possible realization of the chiral MHD turbulence in Weyl metals was also discussed \cite{Galitski}.

One realization of chiral matter in astrophysical systems is the lepton matter in core-collapse supernovae,
where the chirality imbalance of leptons is generated through the electron capture process that involves
only left-handed leptons \cite{Ohnishi:2014uea,Yamamoto:2015gzz}, 
\begin{equation}
\label{weak}
{\rm p} + {\rm e}_{\rm L}^- \rightarrow {\rm n} + \nu^{\rm e}_{\rm L}.
\end{equation} 
Although some of the chirality imbalance of electrons is erased by the finite electron mass 
\cite{Grabowska:2014efa,Kaplan:2016drz}, not all the imbalance is washed out. In particular, production 
of chirality imbalance is more effective as the temperature becomes higher \cite{Sigl:2015xva}
(see also Ref.~\cite{Dvornikov:2014uza} for another possible scenario). 
An alternative mechanism is that a fluid helicity generated by the CVE of the neutrinos {\it effectively} 
plays the role of the chirality imbalance of electrons, and leads to the analog of CME for electrons 
even without such a chirality imbalance \cite{Yamamoto:2015gzz}.

In this paper, we perform the three-dimensional (3D) numerical simulations of fully nonlinear chiral MHD 
equations for the high-density charged chiral matter at the core of a core-collapse supernova and study
the properties of the chiral MHD turbulence. 
As a starting point of numerical simulations and for simplicity, we focus on the CME and chiral anomaly,
but we ignore the CVE, fluid helicity, and cross-helicity, as well as the contributions of the chiral transport 
in the neutrino matter in this paper.
In this sense, our computations here should not be taken as a quantitative prediction.
Rather, our motivation here is to show qualitatively new features due to the chiral effects that have 
been so far disregarded in the context of core-collapse supernovae. 
In fact, we observe the inverse cascade of the magnetic energy and the fluid kinetic energy due to 
the chiral effects in the high-density matter at the supernova core, 
similarly to the high-temperature electroweak plasmas in the early Universe.

This behavior is to be contrasted with conventional multidimensional neutrino radiation-hydrodynamic
simulations for core-collapse supernovae (see Refs.~\cite{Janka:2016fox,Radice:2017kmj} for reviews). 
In most of the 3D supernova models, the direct cascade of turbulent flows in the postshock region is
dominant over the inverse cascade \cite{Hanke:2011jf,Takiwaki:2013cqa,Radice:2017kmj}.
On the other hand, the dominance of the inverse cascade over the direct cascade (leading to a formation
of large-scale flow) has been often observed in axisymmetric (2D) models. This large-scale flow may
account for the vigor explosion found in the 2D models \cite{Janka:2016fox} though the detailed mechanism
is under discussion \cite{kazenori2018}. 
The qualitative difference of our 3D turbulent behaviors from these previous results can be understood 
from the difference of the conservation laws between the two: while the conventional hydrodynamic theory 
respects the conservation of the energy alone, the chiral MHD respects the conservations of not only 
the energy but also a {\it nonzero helicity}.%
\footnote{In the case of 2D fluids, the conservation of {\it enstrophy}, in addition to the conservation of energy, 
leads to the inverse energy cascade \cite{Kraichnan1967}. Although the conservation of enstrophy is absent in three dimensions, 
the conservation of a {\it nonzero helicity} that is specific in three dimensions can lead to the inverse cascade, 
as we will explicitly show in this paper.}
Our results suggest that the chiral effects can reverse the turbulent cascade direction from direct to inverse cascade, 
which may be relevant to the mechanism of supernova explosions.

This paper is organized as follows. In Sec.~\ref{sec:ChMHD}, we formulate the chiral MHD equations 
for protons and electrons with a chirality imbalance in the supernova core. 
In Sec.~\ref{sec:numerics}, we provide the numerical results of the 3D chiral MHD turbulence. 
Sections \ref{sec:discussion} and \ref{sec:conclusion} are devoted to the discussion and conclusion,
respectively. Throughout the paper, we use the units $\hbar = c = e =1$ unless otherwise stated.

\section{Chiral MHD in the supernova core}
\label{sec:ChMHD}

\subsection{Chiral MHD equations}
Here we generalize the conventional MHD by including the CME in the presence of a chirality imbalance 
of electrons generated by the process (\ref{weak}) in the supernova core. The chirality imbalance of 
electrons is characterized by the chiral chemical potential $\mu_{\rm A} \equiv (\mu_{\rm R} - \mu_{\rm L})/2$ 
(or the axial charge density $n_{\rm A}$ defined below), where $\mu_{\rm R, L}$ is the 
chemical potential of the right- or left-handed electron.
Alternatively, the fluid helicity produced by the CVE of neutrinos can be regarded as an 
effective chiral chemical potential $\mu_{\rm A}$.
The precise value of $\mu_{\rm A}$ (including the effective one) in the supernova core is determined by 
the microscopic process (\ref{weak}), the chirality flipping due to the finite electron mass \cite{Grabowska:2014efa}, 
and nonlinear chiral MHD evolutions and is beyond the scope of the present paper. 
In this paper, we will treat $n_{\rm A}$ as a free parameter instead, and we will study the behaviors 
of the chiral MHD turbulence with nonzero $n_{\rm A}$. 

We start from relativistic continuity and momentum equations for the proton with the mass $M$ and electron with the mass $m$ 
(e.g., Ref.~\cite{Balsara:2016lmo}), 
\begin{eqnarray}
  &&\partial_t (\gamma_{\rm p} \rho_{\rm p}) + \bm \nabla\cdot (\rho_{\rm p}\gamma_{\rm p} \mbox{\boldmath $v$}_{\rm p}) = 0 \;, \label{eq:rho_p}\\
  &&\partial_t (\gamma_{\rm e} \rho_{\rm e}) + \bm \nabla\cdot (\rho_{\rm e}\gamma_{\rm e} \mbox{\boldmath $v$}_{\rm e}) = 0 \;, \label{eq:rho_e}\\
  &&\partial_t (\rho_{\rm p} h_{\rm p} \gamma_{\rm p}^2 \mbox{\boldmath $v$}_{\rm p}) + \bm \nabla \cdot (\rho_{\rm p} h_{\rm p}\gamma_{\rm p}^2 \mbox{\boldmath $v$}_{\rm p}\mbox{\boldmath $v$}_{\rm p}) \nonumber \\
&&  =  - {\bm \nabla}P_{\rm p} + \gamma_{\rm p} n_{\rm p} \mbox{\boldmath $E$} + \mbox{\boldmath $J$}_{\rm p} \times \mbox{\boldmath $B$} + \mbox{\boldmath $F$}_{\rm pe} \;, \label{eq:v_p}\\
  &&\partial_t (\rho_{\rm e} h_{\rm e} \gamma_{\rm e}^2 \mbox{\boldmath $v$}_{\rm e}) + \bm \nabla \cdot (\rho_{\rm e} h_{\rm e}\gamma_{\rm e}^2 \mbox{\boldmath $v$}_{\rm e}\mbox{\boldmath $v$}_{\rm e}) \nonumber \\
&&  = -{\bm \nabla}P_{\rm e} - \gamma_{\rm e} n_{\rm e} \mbox{\boldmath $E$} + \mbox{\boldmath $J$}_{\rm e} \times \mbox{\boldmath $B$} + \mbox{\boldmath $F$}_{\rm ep}\;, \label{eq:v_e}
\end{eqnarray}
where $\rho$ is the mass density, \mbox{\boldmath $v$} is the velocity, $\gamma = 1/\sqrt{1-{\bm v}^2}$ is the Lorentz factor, $P$ is the pressure,
$\mbox{\boldmath $E$}$ is the electric field, $\mbox{\boldmath $B$}$ is the magnetic field, \mbox{\boldmath $J$} is the electric current density, 
and $h = 1 + P/\rho + \epsilon$ is the specific enthalpy with $\epsilon$ being the specific internal energy.%
\footnote{Note that the energy density $\varepsilon$ is related to the specific internal energy $\epsilon$ via $\varepsilon = \rho(1+\epsilon)$.}
Subscripts ``p'' and ``e'' denote physical variables of protons and electrons, respectively.
$\mbox{\boldmath $F$}_{\rm pe}$ and $\mbox{\boldmath $F$}_{\rm ep}$ represent the friction force between two fluid species and satisfy   
$\mbox{\boldmath $F$}_{\rm pe} = - \mbox{\boldmath $F$}_{\rm ep}$.
The expressions of ${\bm J}_{\rm p}$ and ${\bm J}_{\rm e}$ are given by 
\begin{eqnarray}
 \mbox{\boldmath $J$}_{\rm p} & =& \gamma_{\rm p} n_{\rm p} \mbox{\boldmath $v$}_{\rm p} \;, \label{eq:J_p} \\
  \mbox{\boldmath $J$}_{\rm e} & =& -\gamma_{\rm e} n_{\rm e} \mbox{\boldmath $v$}_{\rm e} + \xi_{B}\mbox{\boldmath $B$} \label{eq:J_e} \;,
\end{eqnarray}
where $n$ is the number density. The second term of the rhs in Eq.~(\ref{eq:J_e}) is the CME, which arises due to the chirality imbalance of electrons.

For a single right-handed electron with the chemical potential $\mu_{\rm R}$, the chiral magnetic conductivity $\xi_B$ in the Landau frame reads 
\cite{Son:2009tf,Neiman:2010zi,Landsteiner:2011cp},%
\footnote{We note that there is an ambiguity on the choice of the frame in relativistic hydrodynamics. 
In the frame where the CME takes the familiar form of the electric current, ${\bm j}_{\rm CME} = \mu_{\rm R}{\bm B}/(4\pi^2)$, 
the CME also contributes to the energy-momentum tensor, e.g., $T^{0i}_{\rm CME} = T^{i0}_{\rm CME} = \mu_{\rm R}B^i/(8\pi^2)$ \cite{Son:2009tf,Neiman:2010zi,Landsteiner:2011cp},
which would change the momentum equation (\ref{eq:v}) below. 
(Such contributions seem to be missed in Refs.~\cite{Schober:2017cdw,Brandenburg:2017rcb,Schober:2018ojn}.)
Following Ref.~\cite{Yamamoto:2015gzz}, we take here the Landau frame such that the CME contributes to ${\bm j}$, but not to $T^{\mu \nu}$.}
\begin{equation}
\xi_B = \frac{\mu_{\rm R}}{4\pi^2}\left( 1 - \frac{1}{2}\frac{n_{\rm R}\mu_{\rm R}}{\varepsilon + P} \right) - \frac{1}{24}\frac{n_{\rm R}T^2}{\varepsilon + P} \simeq \frac{\mu_{\rm R}}{8\pi^2}\;.
\label{eq:CME}
\end{equation}
where $n_{\rm R}$ and $\varepsilon$ are the charge and energy densities of right-handed electrons. 
In the last line, we used $\mu_{\rm R} \gg T$ (expected in the supernova core) and the thermodynamic relation $\varepsilon + P \simeq \mu_{\rm R} n_{\rm R}$.
In the system with right- and left-handed electrons, the equation above is modified to the form with 
replacing $\mu_{\rm R}$ with $\mu_{\rm R} - \mu_{\rm L}$. 
Since the chemical potential is linked to the charge density via the relation $n_i = \mu_i^3/(6\pi^2)\ (i={\rm R,L})$ 
for a noninteracting relativistic Fermi gas (which we assume for simplicity) with $\mu \gg T$,%
\footnote{In general, the charge density of a noninteracting relativistic Fermi gas is given, 
as functions of the chemical potential and the temperature, by
  \begin{equation}
    n = \frac{\mu^3}{6\pi^2} + \frac{\mu T^2}{6} \;. \nonumber
  \end{equation}
  In the limit $\mu \gg T$ which is the regime of our interest, the first term on the rhs becomes dominant. In contrast, for $\mu \ll T$ relevant to the
  early Universe studied in Refs.~\cite{Brandenburg:2017rcb,Schober:2018ojn}, the second term mainly determines the charge density.}
the chiral magnetic conductivity is given by
\begin{eqnarray}
\! \! \! \!  \xi_B & = & \frac{1}{8\pi^2}\left( \mu_{\rm R} - \mu_{\rm L}\right) \nonumber \\
  & = & \frac{1}{8} \! \left(\frac{3}{\pi^4} \right)^{\! \!1/3} \! \left[ \left(n_{\rm e}+n_{\rm A}\right)^{1/3} \! - \! \left(n_{\rm e}-n_{\rm A} \right)^{1/3}\right]\,, \label{eq:xi_B}
\end{eqnarray}
where $n_{\rm A} = n_{\rm R} - n_{\rm L}$ is the axial charge density and 
$n_{\rm e} = n_{\rm R} + n_{\rm L}$ is the total charge density of electrons, which can be replaced approximately
by $\rho/M$ because of the charge neutrality ($n_{\rm e} = n_{\rm p}  = \rho/M$) that we assume in our system.

In the following, we will derive one-fluid chiral MHD equations from two-fluid hydrodynamic equations for protons and electrons above.
We assume that $|\mbox{\boldmath $v$}_{\rm p}| \ll 1$ and $|\mbox{\boldmath $v$}_{\rm e}| \ll 1$ and then use
the nonrelativistic approximations $\gamma \approx 1$ and $h \approx 1$, which can be justified in the case of the supernova core. 

First, we derive one-fluid continuity and momentum equations from Eqs.~(\ref{eq:rho_p})--(\ref{eq:J_e}). 
As usual, the continuity equation is obtained from the sum of Eqs.~(\ref{eq:rho_p}) and (\ref{eq:rho_e}) as
\begin{equation}
  \partial_t \rho + \bm \nabla\cdot (\rho \mbox{\boldmath $v$}) = 0 \label{eq:rho}\;, 
\end{equation}
where the one-fluid mass density $\rho$ and velocity $\mbox{\boldmath $v$}$ defined below can be approximated for $M \gg m$ by
\begin{eqnarray}
  \rho & \equiv & \rho_{\rm p} + \rho_{\rm e} = n(M+m) \simeq nM \;, \label{eq:rho_sum}\\
  \mbox{\boldmath $v$} & \equiv & \frac{M\mbox{\boldmath $v$}_{\rm p} + m \mbox{\boldmath $v$}_{\rm e}}{M + m} \simeq \mbox{\boldmath $v$}_{\rm p} \;. \label{eq:v_sum} 
\end{eqnarray}

The momentum equation is obtained from the sum of Eqs.~(\ref{eq:v_p}) and (\ref{eq:v_e}) for $M \gg m$ and $M \gg mh_{\rm e}$ by
\begin{equation}
  \partial_t (\rho \mbox{\boldmath $v$} ) + \bm \nabla \cdot (\rho \mbox{\boldmath $v$}\mbox{\boldmath $v$})
  =  - {\bm \nabla}P + \mbox{\boldmath $J$} \times \mbox{\boldmath $B$} \label{eq:v2} \;. 
\end{equation}
where $ P \equiv P_{\rm p} + P_{\rm e}$ is the total pressure, and $\mbox{\boldmath $J$} $ is the total current density
defined by
\begin{eqnarray}
\! \! \! \! \! \mbox{\boldmath $J$} &\equiv& \mbox{\boldmath $J$}_{\rm p} + \mbox{\boldmath $J$}_{\rm e} \nonumber \\
  &=& [n (\mbox{\boldmath $v$}_{\rm p} - \mbox{\boldmath $v$}_{\rm e})] + [\xi_{B}\mbox{\boldmath $B$}] = \mbox{\boldmath $J$}_{\rm MHD} + \mbox{\boldmath $J$}_{\rm CME} \;,
  \label{eq:J_tot}
\end{eqnarray} 
where $\mbox{\boldmath $J$}_{\rm MHD}$ is the current density due to the velocity difference between positive and negative charges, and $\mbox{\boldmath $J$}_{\rm CME}$
is that due to the CME. By adding the viscous term in Eq.~(\ref{eq:v2}) as usual (see, e.g., Refs.~\cite{KT73,GB05}), we obtain
\begin{equation}
\rho \mathcal{D}_t{\bm v}   = - \bm \nabla P + \mbox{\boldmath $J$} \times {\bm B} + \bm \nabla \cdot {\bm \Pi} \;, \label{eq:v} 
\end{equation}
where $\mathcal{D}_t$ is the Lagrangian time derivative and 
$\Pi_{ij}  =   2\rho \nu S_{ij}$ is the the viscous stress tensor with the viscosity $\nu$ and the strain rate tensor,
\begin{equation}
S_{ij} = \frac{1}{2}\left( \partial_j v_i + \partial_i v_j - \frac{2}{3}\delta_{ij}\partial_i v_i \right) \;.  \label{eq:S} 
\end{equation}
The bulk viscosity is neglected for simplicity in this study.

Let us next derive Ohm's law and the resulting energy and induction equations. 
Multiplying Eq.~(\ref{eq:v_p}) by $m/\rho$ and Eq.~(\ref{eq:v_e}) by $M/\rho$ and taking their difference, we have
\begin{align}
m D_t\left( \frac{\mbox{\boldmath $J$}_{\rm MHD}}{n} \right) + \frac{1}{\rho}(m\bm \nabla P_{\rm p}- M\bm \nabla P_{\rm e})
\nonumber \\
= \mbox{\boldmath $E$} + \frac{1}{\rho}(m\mbox{\boldmath $J$}_{\rm p} - M\mbox{\boldmath $J$}_{\rm e})\times \mbox{\boldmath $B$}
- \eta\mbox{\boldmath $J$}_{\rm MHD} \;, \label{eq:Ohm1} 
\end{align}
where $\eta$ is the resistivity and the canonical relation of $\mbox{\boldmath $F$}_{\rm ep} = \eta n \mbox{\boldmath $J$}_{\rm MHD}$  is used. 
The second term of the rhs is rewritten by 
\begin{equation}
  \frac{1}{\rho}(m\mbox{\boldmath $J$}_{\rm p} - M\mbox{\boldmath $J$}_{\rm e}) \times \mbox{\boldmath $B$} \! = \! \left(\mbox{\boldmath $v$}
 \! - \! \frac{M \! - \! m}{\rho}\mbox{\boldmath $J$}_{\rm MHD} \! - \! \frac{M}{\rho}\mbox{\boldmath $J$}_{\rm CME} \right) \times \mbox{\boldmath $B$} \;. 
\end{equation}
For $M \gg m$ and for a sufficiently long timescale $t \gg 1/\omega_{\rm pe}$ with $\omega_{\rm pe}$ being the plasma frequency of electrons, 
Eq.~(\ref{eq:Ohm1}) becomes
\begin{equation}
  \mbox{\boldmath $E$} + \mbox{\boldmath $v$} \times \mbox{\boldmath $B$}  - \eta\mbox{\boldmath $J$}_{\rm MHD}
  = \frac{1}{n}\mbox{\boldmath $J$} \times \mbox{\boldmath $B$} - \frac{1}{n}\bm \nabla P_{\rm e} \;, \label{eq:Ohm2}
\end{equation}
where the first and second terms on the rhs are the Hall term and the electron pressure term, respectively.
In Eq.~(\ref{eq:Ohm2}), we neglect the electron inertia and the proton pressure by focusing on lower-frequency
motions of the plasma than the electron plasma oscillation due to the local charge separation. This is an essential
difference between our one-fluid chiral MHD system and the original two-fluid description of the chiral plasma.
When ignoring the terms on the rhs for simplicity, the modified Ohm's law including the CME becomes 
\begin{equation}
\mbox{\boldmath $E$} + \mbox{\boldmath $v$} \times \mbox{\boldmath $B$} = \eta (\mbox{\boldmath $J$} - \mbox{\boldmath $J$}_{\rm CME}) \;. \label{eq:Ohm3}
\end{equation}
Note that the Joule-heating term in the internal energy equation can be evaluated from Eq.~(\ref{eq:Ohm3}) by $\mbox{\boldmath $J$}\cdot \mbox{\boldmath $E$}$. 
Hence, the energy equation is given by 
\begin{eqnarray}
  \rho \mathcal{D}_t\epsilon  = - P\bm \nabla\cdot {\bm v}  + 2\rho\nu \bm{S}^2
  + \eta\mbox{\boldmath $J$}\cdot (\mbox{\boldmath $J$} - \mbox{\boldmath $J$}_{\rm CME})\;. \label{eq:E} 
\end{eqnarray}

Furthermore, from Faraday's law $\partial_t \mbox{\boldmath $B$} = - \bm \nabla \times \mbox{\boldmath $E$}$, 
the induction equation modified by the CME is obtained as
\begin{equation}
  \partial_t \mbox{\boldmath $B$} = \bm \nabla \times (\mbox{\boldmath $v$}\times \mbox{\boldmath $B$})
  + \eta \bm \nabla^2 \mbox{\boldmath $B$}  + \eta \bm \nabla \times \left( \xi_{B}\mbox{\boldmath $B$} \right) \;. \label{eq:B}
\end{equation}
Here, Amp\`ere's law $\mbox{\boldmath $J$} = \bm \nabla \times \mbox{\boldmath $B$}$ is used. The last term on the rhs
is the correction due to the CME. 

Finally, the time evolution of $n_{\rm A}$ is given by the chiral anomaly equation \cite{Adler,BellJackiw},
\begin{equation}
\partial_{\mu} J^{\mu}_{\rm A} = \frac{1}{2\pi^2}{\bm E} \cdot {\bm B}\;.
\end{equation}
where $J^{\mu}_{\rm A}$ is the axial 4-current. Using Eq.~(\ref{eq:Ohm2}), this can be rewritten as
\begin{equation}
  \partial_t n_{\rm A} = \frac{\eta}{2\pi^2}\left(\mbox{\boldmath $J$}
  - \mbox{\boldmath $J$}_{\rm CME} \right)\cdot \mbox{\boldmath $B$} \;, \label{eq:n_A}
\end{equation}
where, for our step-by-step strategy, we ignore the advection term, the diffusion term, the so-called chiral separation
effect (CSE) \cite{Son:2004tq,Metlitski:2005pr}, ${\bm J}_{\rm CSE}^{\rm A} = \xi^{\rm A}_B {\bm B}$ 
with $\xi^{\rm A}_{B}$ being the transport coefficient 
and the cross-helicity $J_{\rm CSE}^{{\rm A}0} =  \xi^{\rm A}_B {\bm v} \cdot {\bm B}$, for simplicity. 
Note that, under this simplification, Eq.~(\ref{eq:n_A}) can be understood as the conservation of helicity
(fermion helicity plus magnetic helicity, but without cross helicity and fluid helicity) \cite{Yamamoto:2015gzz}; 
see also the remark below.

For an equation of state (EOS), we adopt the ideal gas law, 
\begin{equation}
\label{eq:eos}
P = (\Gamma -1)\rho\epsilon \;,
\end{equation}
for simplicity, where $\Gamma = 5/3$ in the adiabatic index. Then, we can close the system. 
The set of Eqs. (\ref{eq:xi_B}), (\ref{eq:rho}), (\ref{eq:v}), (\ref{eq:E}), (\ref{eq:B}), and (\ref{eq:n_A})
coupled with the EOS (\ref{eq:eos}) is solved simultaneously in our simulation. 

Before proceeding further, we comment on several simplifications of our formulation in this paper.
Here and below, we focus on the CME and chiral anomaly, but for simplicity, 
we ignore the CVE, ${\bm J}_{\rm CVE} = \xi_{\omega}{\bm \omega}$, 
the CSE expressed by ${\bm J}_{\rm CSE}^{\rm A} = \xi^{\rm A}_B {\bm B} + \xi^{\rm A}_{\omega} {\bm \omega}$, 
and the other types of helicity (fluid helicity and cross-helicity).
Here, $\mbox{\boldmath $\omega$} \equiv {\bm \nabla} \times {\bm v}$ is the fluid vorticity, 
and $\xi_{\omega}$ is the chiral vortical conductivity.
In particular, we ignore the contributions of the chiral effects of neutrinos.
Incorporating the CVE and CSE is necessary to ensure the conservation of total helicity 
(summation of the fermion helicity, magnetic helicity, fluid helicity, and cross-helicity), 
which would be an important question to be studied in the future; 
see Refs.~\cite{Yamamoto:2015gzz,Avdoshkin:2014gpa} for such a generic conservation law of helicity.
The importance of the CVE in the chiral MHD turbulence will briefly be discussed in Sec.~\ref{sec:CVE}.

\subsection{Chiral plasma instability}
\label{sec:CPI}

\begin{figure*}[htbp]
\begin{center}
\vspace{-20pt}
\scalebox{0.75}{{\includegraphics{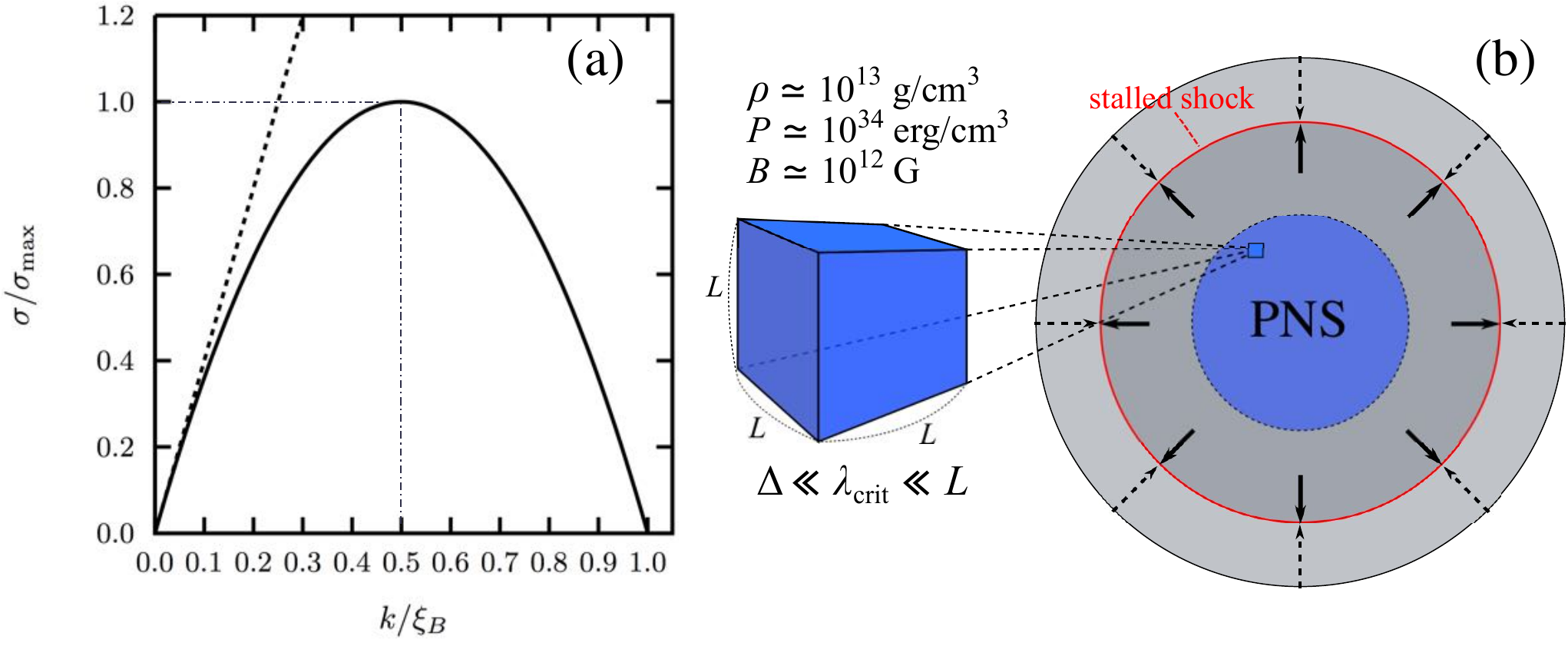}}} 
\caption{(a) Dispersion relation of the CPI. 
  (b) Setup for our local box simulation. The global structure of the supernova core (right)
  and extracted local Cartesian patch (left). The box size $L$ is chosen so as to resolve
  $\lambda_{\rm crit}$. }
\label{CPI}
\end{center}
\end{figure*}

To build up the simulation model, we should bear in mind the driving mechanism of the chiral MHD turbulence.
The presence of the CME induces the amplification of the magnetic field due to the chiral plasma instability (CPI); 
see, e.g., Refs.~\cite{Akamatsu:2013pjd,Ohnishi:2014uea,Yamamoto:2015gzz} in the context of neutron stars and supernovae. 
By inserting the perturbation $\delta \mbox{\boldmath $B$} \propto \exp(i{\bm k}\cdot {\bm x} + \sigma t)$ 
into the induction equation (\ref{eq:B}) around the stationary uniform background $\mbox{\boldmath $v$} = \mbox{\boldmath $B$} = {\bm 0}$, 
we obtain the dispersion equation for the CPI,
\begin{equation}
\sigma = \eta\xi_B k - \eta k^2\;, 
\label{eq:CPI}
\end{equation}
where $k = |{\bm k}|$ is the wave number. Shown by the solid line in Fig.~\ref{CPI}(a) is the real part of $\sigma = \sigma(k)$ that
characterizes the growth rate of the CPI. 
The dashed line denotes the linear dispersion relation neglecting the $k^2$ term in Eq.~(\ref{eq:CPI}).
The vertical and horizontal axes are normalized by the maximum growth rate $\sigma_{\rm max} = \eta\xi_B^2/4$ and $\xi_B$, respectively.
The CPI grows only in the low-wave number regime,
\begin{equation}
k < k_{\rm crit} = \xi_B \;, \label{eq:k_crit}
\end{equation}  
and becomes maximum when $k = \xi_B/2$, indicating that the onset of the CPI is due solely to the presence of the chirality imbalance.
While the growth rate of the CPI becomes larger with the increase of $\eta$, the magnetic diffusion term plays a role in, as usual,
suppressing the modes in the high-wave number regime $k > k_{\rm crit}$ (see the dashed line). The typical wavelength of the CPI becomes
longer with the decrease of $\xi_B$, suggesting the tendency toward inverse energy cascade under the situation in which $\xi_B$ decreases
as a function of time. 

It is worth noting here that the CME is somewhat similar to the $\alpha$ effect in the mean-field dynamo theory \cite{MFT78,PK79,KR80}. 
When dividing the variables of the induction equation into the ensemble-averaged values and fluctuating components, the $\alpha$ effect
appears in the turbulent electromotive force as an induction term. It is a consequence of the forced symmetry breaking in rotating
astrophysical bodies, such as the Sun, stars, and accretion disks, and has been widely studied as an origin of the large-scale magnetic
field commonly observed in these objects \cite{Brandenburg:2004jv,brandenburg18,masada+14b,masada+16}. 
On the other hand, the CME is a pure quantum effect that originates from the explicit parity symmetry breaking by the chirality of fermions. 
Although one can expect common traits in the macroscopic hydrodynamic evolutions between them 
(see Refs.~\cite{Rogachevskii:2017uyc,Brandenburg:2017rcb,Schober:2018ojn,dvornikov+17} for the chiral MHD turbulence in the early Universe),
there is also a difference: the total helicity is vanishing for the $\alpha$ effect in the nonchiral matter, whereas this is not the case in
chiral matter. In particular, a nonzero magnetic helicity can be generated {\it globally} in chiral matter.

\section{Numerical results}
\label{sec:numerics}
\subsection{Simulation setup \label{RS1}}
\begin{table*}[htbp]	
\begin{tabular}{c c c c c c c c c} \hline\hline
  & $N^3 $ & $\eta$ & $L$ & $n_{\rm A}$ & $\xi_{B,{\rm ini}}$  & $\tau_{\rm CPI}$ & $B_{\rm sat}$ & $\xi_{B,{\rm sat}}/\xi_{B,{\rm ini}}$ \\ \hline\hline
Model 1  \ &$256^3$ &$ 100.0 $ & $2\times 10^4 $ & $0.1$ & $4.2 \times 10^{-3}$ & $2.3 \times 10^3$ & $2.5 \times 10^{-2}$ & $0.077$ \\ 
Model 2  \ &$128^3$ &$ 100.0 $ & $2\times 10^4 $ & $0.1$ & $4.2 \times 10^{-3}$ & $2.3 \times 10^3$ & $2.3 \times 10^{-2}$ & $0.075$ \\ 
Model 3  \ &$64^3 $ &$ 100.0 $ & $2\times 10^4 $ & $0.1$ & $4.2 \times 10^{-3}$ & $2.3 \times 10^3$ & $1.4 \times 10^{-2}$ & $0.077$ \\ 
Model 4  \ &$128^3$ &$ 10.0  $ & $2\times 10^4 $ & $0.1$ & $4.2 \times 10^{-3}$ & $2.3 \times 10^4$ & $2.5 \times 10^{-2}$ & $0.078$ \\ 
Model 5  \ &$128^3$ &$ 1.0   $ & $2\times 10^4 $ & $0.1$ & $4.2 \times 10^{-3}$ & $2.3 \times 10^5$ & $2.7 \times 10^{-2}$ & $0.077$ \\ 
Model 6  \ &$128^3$ &$ 1.0   $ & $1\times 10^5 $ & $0.1$ & $4.2 \times 10^{-3}$ & $2.3 \times 10^5$ & $1.1 \times 10^{-2}$ & $0.024$ \\ 
Model 7  \ &$64^3$ &$ 1.0   $ & $1\times 10^4 $ & $0.1$ & $4.2 \times 10^{-3}$ & $2.3 \times 10^5$ & $3.6 \times 10^{-2}$ & $0.15$ \\ 
Model 8  \ &$32^3$ &$ 1.0   $ & $4\times 10^3 $ & $0.1$ & $4.2 \times 10^{-3}$ & $2.3 \times 10^5$ & $4.9 \times 10^{-2}$ & $0.39$ \\ 
Model 9  \ &$128^3$ &$ 1.0   $ & $4\times 10^3 $ & $0.416$ & $2.1 \times 10^{-2}$ & $9.2 \times 10^3$ & $1.2 \times 10^{-1}$ & $0.075$ \\ 
Model 10 \ &$128^3$ &$ 1.0   $ & $1\times 10^5 $ & $0.020$ & $8.4 \times 10^{-4}$ & $5.7 \times 10^6$ & $5.5 \times 10^{-3}$ & $0.078$ \\ 
\hline\hline
\end{tabular}
\caption{Summary of the simulation runs. 
Two diagnostic quantities $B_{\rm sat}$ and $\xi_{B,{\rm sat}}$ are defined by $B_{\rm sat} \equiv \langle \mbox{$B$}^2(t_{\rm sat}) \rangle^{1/2}$ and
$\xi_{B,{\rm sat}} \equiv \langle \xi_B (t_{\rm sat}) \rangle$, where $t_{\rm sat}$ is the time when the system reaches saturation. 
See the text for further explanations.} 
\label{setup}
\end{table*}

We perform a series of 3D simulations by adopting a local Cartesian model, which zooms in on a small patch of the proto-neutron star (PNS),
with the cubic periodic box. See Fig.~\ref{CPI}(b) for the schematic view of our numerical model. In most of our simulations, the size of the simulation
domain $L$ and the grid size $\Delta \equiv L/N$ ($N$ is the number of numerical grids) are determined so as to resolve the critical wavelength 
of the CPI defined by $\lambda_{\rm crit}\equiv 2\pi/\xi_B$, that is,  
\begin{equation}
  \Delta \ll \lambda_{\rm crit} \ll L \;. \label{condition}
\end{equation}
We will discuss how the resolution and the box size of the simulation model affect the behaviors of the chiral MHD turbulence
in Secs.~\ref{RS3} and ~\ref{RS5}. 

The MHD equations are solved by the second-order Godunov-type finite-difference scheme that 
employs an approximate MHD Riemann solver \cite{sano+99,masada+15}. The magnetic field evolves with the Consistent
Method of Characteristics-Constrained Transport (MoC-CT) scheme with including the CME as a part of the electromotive force
(see Refs.~\cite{evans+88,clarke96} for the MoC-CT method).
The chirality imbalance is updated according to Eq.~(\ref{eq:n_A}) straightforwardly with the MHD variables. 

All the numerical models have the same initial density and pressure of $\rho = 5.0$ and $P = 1.0$ in the unit of $100\ {\rm MeV} = 1$, 
which are equivalent to $\rho \simeq 10^{13}\ {\rm g/cm^3}$ and $P \simeq 10^{34}\ {\rm erg/cm^3}$ of the typical PNS
(see, e.g., Ref.~\cite{Kotake:2012iv}).
For the fiducial run, we adopt the uniform axial charge density $n_{\rm A} = 0.1\ (\equiv n_{\rm A0})$, which corresponds
to $\xi_{B} = 4.2 \times 10^{-3}\ (\equiv \xi_{B0})$,
the uniform viscosity $\nu = 0.1\ (\equiv \nu_0)$, the uniform resistivity $\eta = 100.0\ (\equiv \eta_0)$, and a resolution of $N^3 = 256^3$ grid
points. Since $\lambda_{\rm crit}$ for these values of the parameters is estimated as $\lambda_{\rm crit} = 1.5\times 10^3$, 
we choose $L = 2\times 10^4 (\equiv L_0)$ to satisfy the condition (\ref{condition}). 

As stated in Sec.~\ref{sec:introduction}, a few physical processes are relevant for the origin of the 
finite axial charge density $n_{\rm A}$.
One scenario is the chiral imbalance produced by the electron capture process \eqref{weak} before
the neutrino trapping ($\rho_c \lesssim 10^{12} {\rm g/cm^3}$ with $\rho_{\rm c}$ the central density of stars).
After the neutrino trapping, the electron capture process slowly proceeds since
the inverse process blocks the production of the imbalance and 
diffusion process of the neutrino controls the net rate. 
In an alternative scenario, the fluid helicity of the trapped neutrino also plays the role of $n_{\rm A}$.
This fluid helicity could be induced by the rotation or the convection of the star.
In this study, we change $n_{\rm A}$ parametrically since we cannot treat these effects in a self-consistent 
manner that requires a global neutrino radiation-hydrodynamics simulation. 

The resistivity $\eta$ of moderately degenerate electrons is expected to be of the order of $0.1$--$1.0$ (under
the scaling of $100\ {\rm MeV} = 1$) in the PNS (see, e.g., Ref.~\cite{Rossi:2008xw}). Since $\eta$ chosen in our
fiducial model is larger than the actual value, the dependence of the chiral MHD turbulence on $\eta$ is studied
in Sec.~\ref{RS4}. In contrast, the viscosity $\nu$ due to electrons is expected to be $\mathcal{O}(10^{-2})$ and
is compatible with the value chosen in the fiducial run. The dependence of the chiral MHD turbulence on the magnetic
Prandtl number ($\equiv \nu/\eta$) is beyond the scope of this study but a target of our future work. 

In addition to the fiducial run, we simulate a number of models with varying model parameters, such as $N$,
$\eta$, $L$, and $n_{\rm A}$, to study their impacts on the chiral MHD turbulence. The parameters adopted in our
simulation models are listed in Table~\ref{setup} with a few diagnostic quantities. A random small ``seed'' magnetic
field with the amplitude $|\delta {\bm B}| < 0.01$, which gives the maximum value $\simeq 10^{12}\ {\rm G}$
in the cgs unit, is introduced into the initial stationary state with $|{\bm v}| = 0.0$.

\subsection{Fiducial run \label{RS2}} 
\begin{figure}[htbp]
\begin{center}
\scalebox{0.42}{{\includegraphics{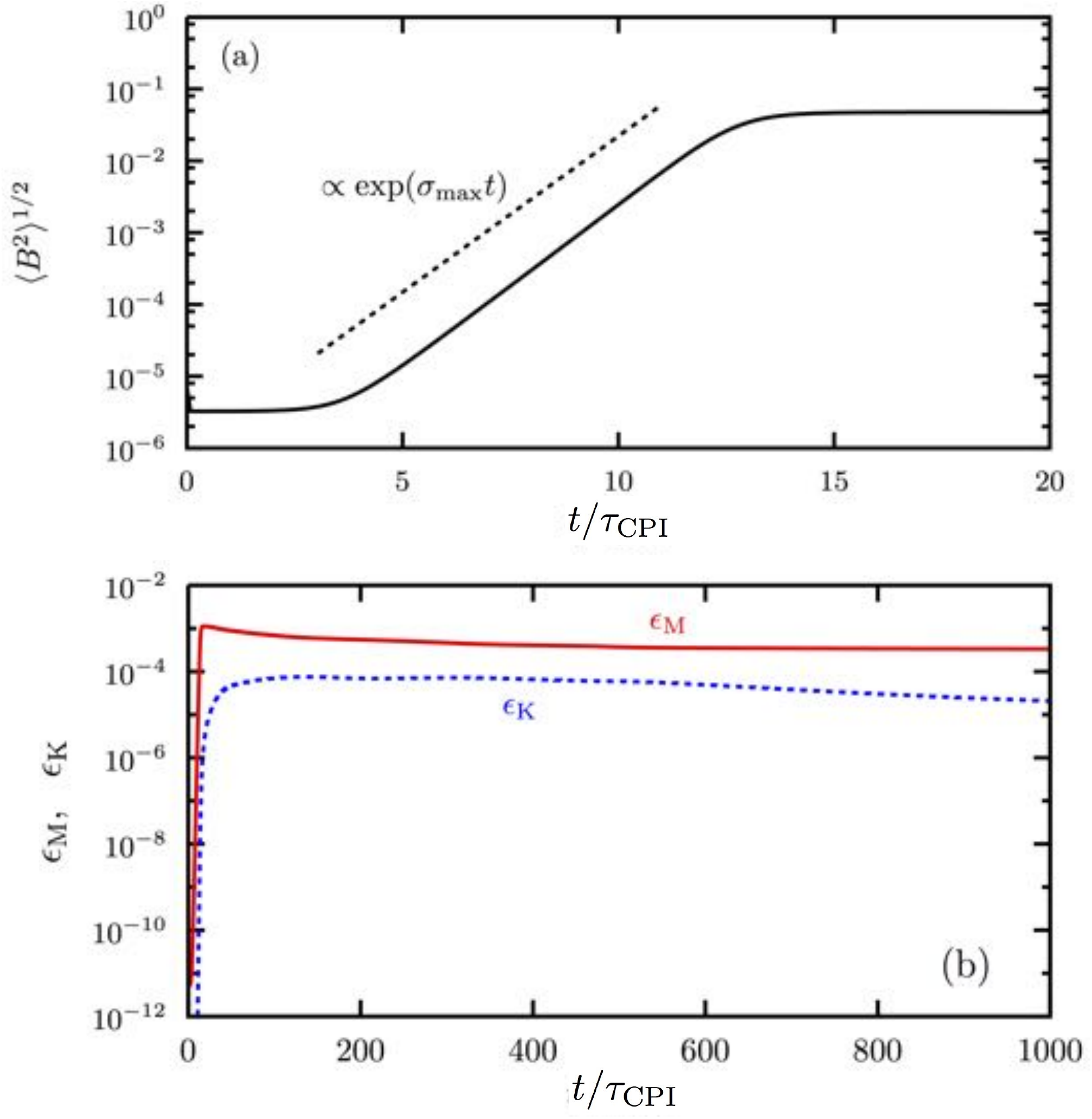}}}  
\caption{Temporal evolutions of $\langle B^2 \rangle^{1/2}$ at the early exponential-growth stage [panel (a)]
and $\epsilon_{\rm M}$ (red solid) and $\epsilon_{\rm K}$ (blue dashed) [panel (b)] for the fiducial run.}
\label{B}
\end{center}
\end{figure}
The basic property in the evolution and saturation of the chiral MHD turbulence is illustrated with the fiducial
model (model~1) as an example. We first define the strength of the mean magnetic field by 
\begin{equation} 
  \langle B^2 \rangle^{1/2} \equiv \left(\frac{1}{V} \int B^2 {\rm d}^3 {\bm x} \right)^{\! 1/2}\,,
  \quad V \equiv \int {\rm d}^3 {\bm x}\,, \label{eq:B_ave}
\end{equation}
where $B \equiv |{\bm B}|$ is the magnitude of the magnetic field and the angular brackets denote the volume average. 
Similarly, the mean magnetic and kinetic energies are defined by 
\begin{equation} 
\epsilon_{\rm M} \equiv \frac{1}{4\pi} \langle B^2 \rangle\,, \qquad 
\epsilon_{\rm K} \equiv \frac{1}{2} \langle \rho v^2 \rangle\,,
\end{equation}
respectively, where $v \equiv |{\bm v}|$ is the magnitude of the flow velocity.
Figure~\ref{B}(a) shows the temporal evolution of $\langle B^2 \rangle^{1/2}$ at the early evolutionary stage. In
addition, the evolutions of $\epsilon_{\rm M}$ (red solid) and $\epsilon_{\rm K}$ (blue dashed) are also
depicted in Fig.~\ref{B}(b). The horizontal axes are normalized by the growth time of the most growing mode of
the CPI defined by 
\begin{equation}
\tau_{\rm CPI} \equiv \sigma_{\rm max}^{-1} = 4(\eta\xi_{B,{\rm ini}}^2)^{-1} \;, \label{eq:t_CPI}
\end{equation}
where $\xi_{B,{\rm ini}}$ is the initial value of $\xi_B$. For the fiducial model, $\tau_{\rm CPI}$ is evaluated
as $2.3\times 10^3$. The dashed line in the panel~(a) is a reference slope proportional to $\exp(\sigma_{\rm max} t)$. 

\begin{figure}[htbp]
\begin{center}
\scalebox{0.4}{{\includegraphics{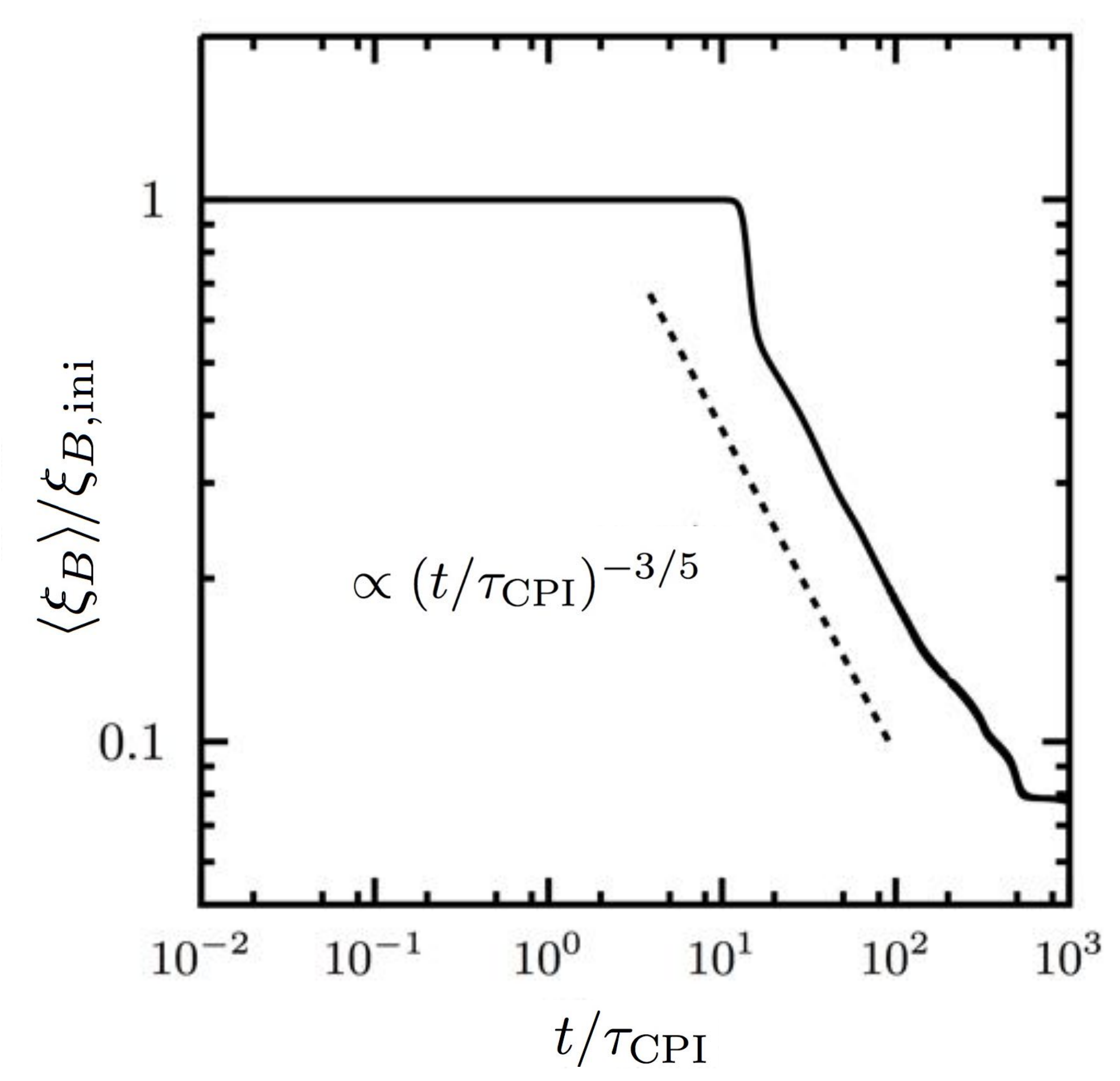}}}  
\caption{Temporal evolution of $\langle \xi_B \rangle$ normalized by $\xi_{B,{\rm ini}}$. 
The dashed line is a reference slope proportional to $(t/\tau_{\rm CPI})^{-3/5}$.} 
\label{xi_B}
\end{center}
\end{figure}
\begin{figure*}[htbp]
\begin{center}
\scalebox{0.8}{{\includegraphics{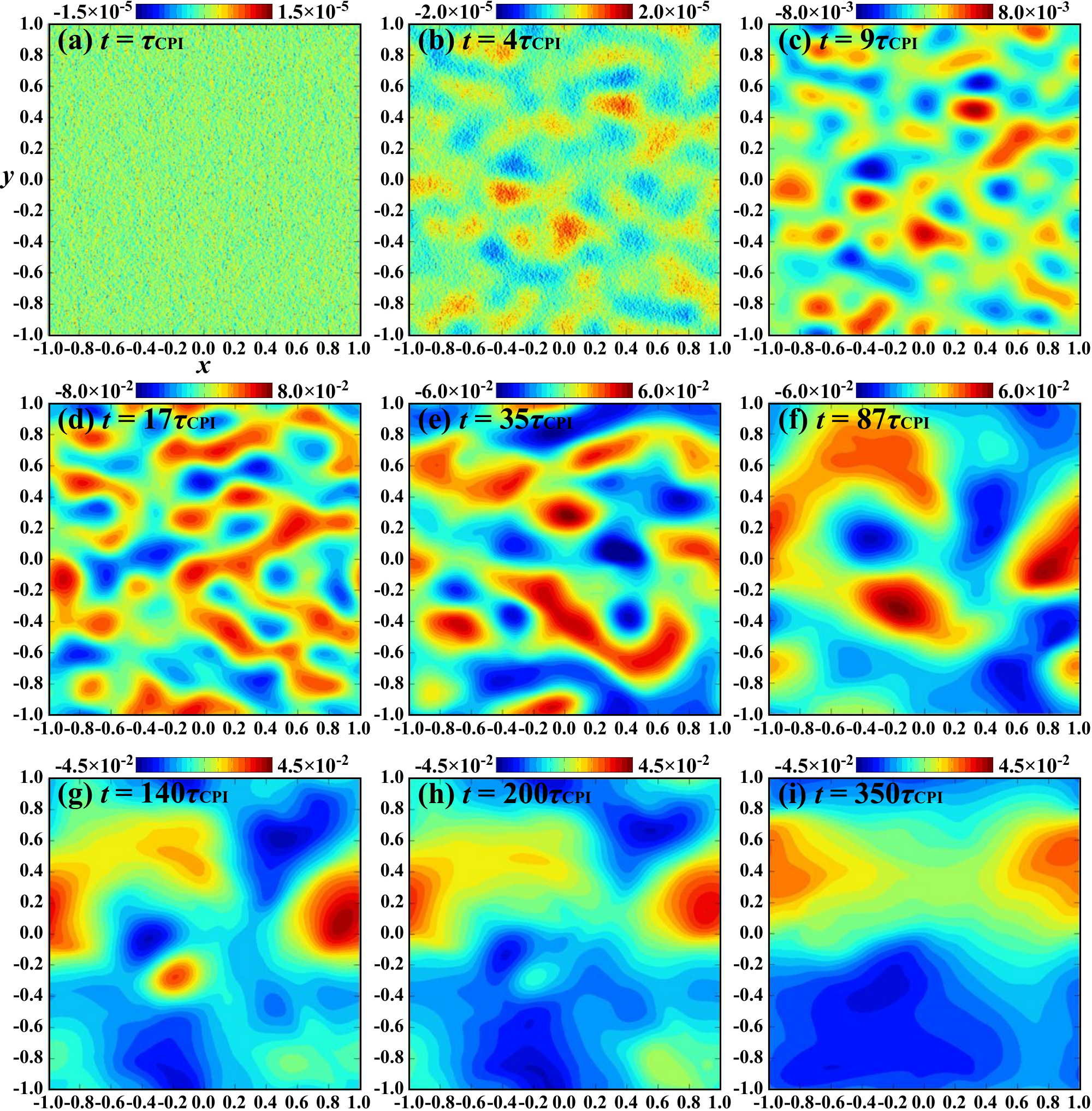}}}
\caption{A series of snapshots of the distribution of $B_x$ on the $x$--$y$ cutting plane at $z = 0$. 
The vertical and horizontal axes are both normalized by $L/2$.}
\label{B_x}
\end{center}
\end{figure*}
\begin{figure*}[htbp]
\begin{center}
\scalebox{0.8}{{\includegraphics{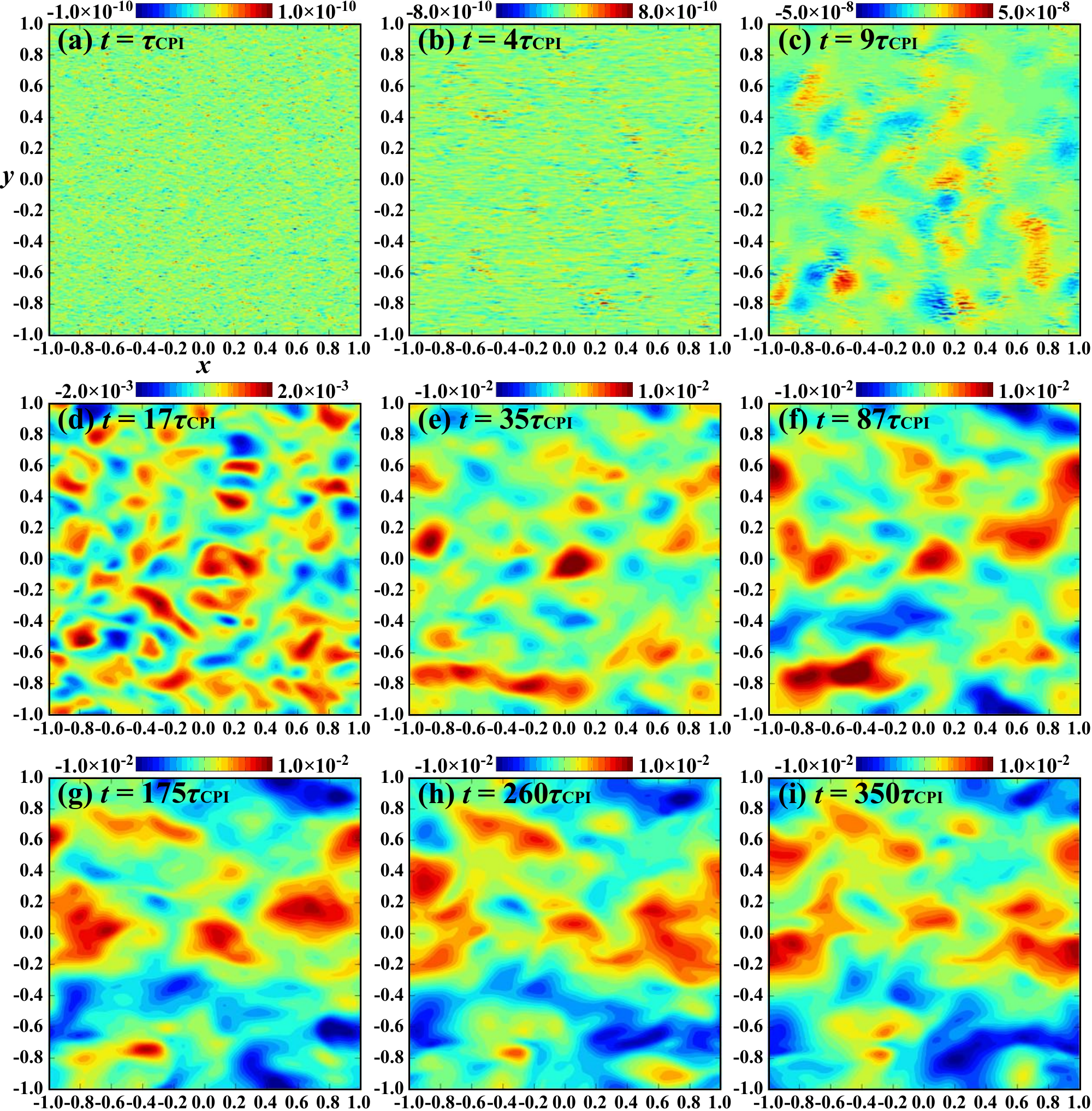}}}  
\caption{A series of snapshots of the distribution of $v_x$ on the $x$--$y$ cutting plane at $z = 0$.
The vertical and horizontal axes are both normalized by $L/2$.} 
\label{v_x}
\end{center}
\end{figure*}
\begin{figure}[htbp]
\begin{center}
\scalebox{0.4}{{\includegraphics{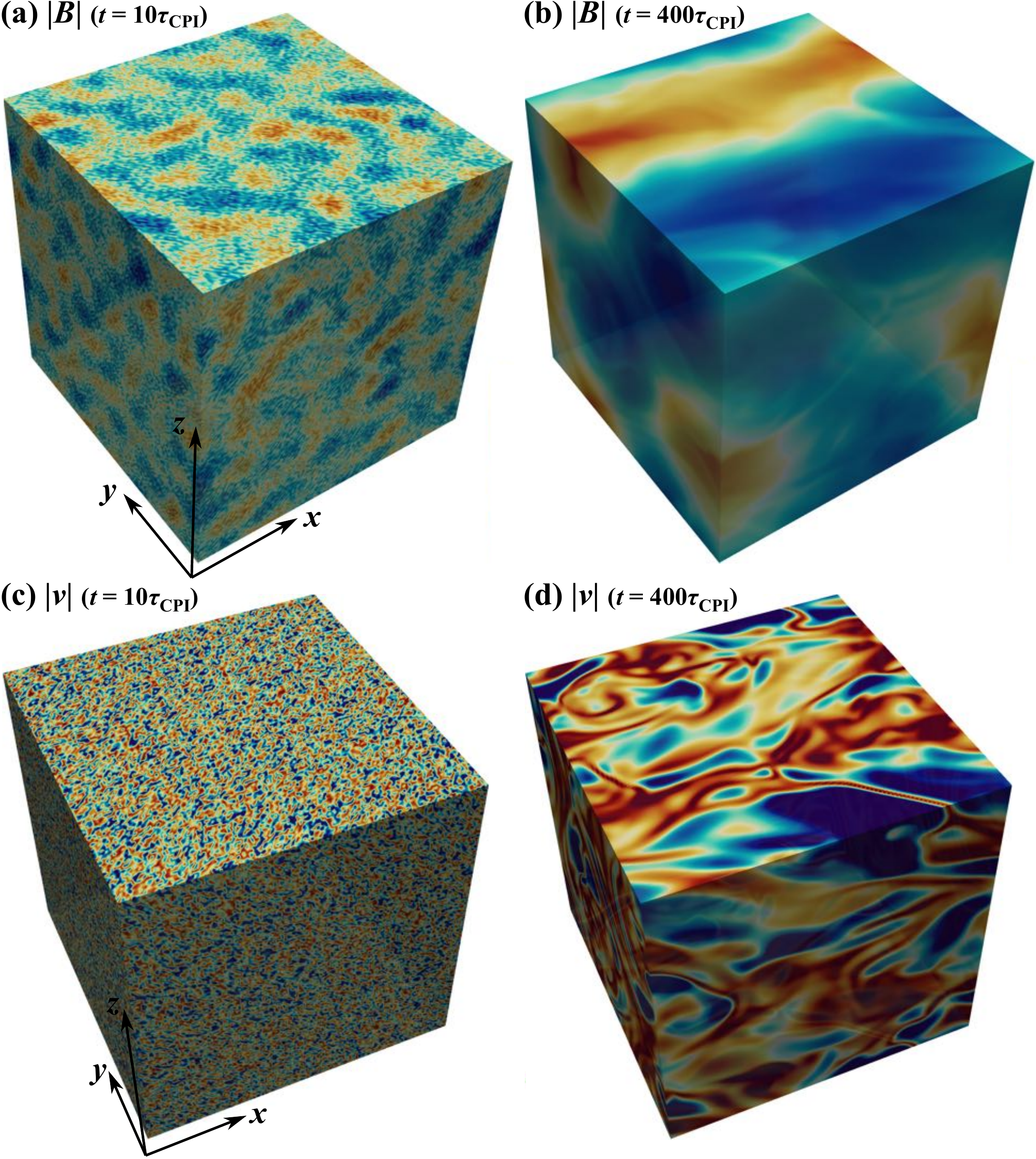}}}  
\caption{3D visualization of $B$ [panels~(a) and~(b)] and $v$ [panels~(c) and~(d)] at the early evolutionary
  stage ($t=10\tau_{\rm CPI}$) and fully nonlinear stage ($t=400\tau_{\rm CPI}$), respectively. The red and blue
  tones denote the lower and higher magnitudes of the fields. } 
\label{3D}
\end{center}
\end{figure}
\begin{figure*}[htbp]
\begin{center}
\scalebox{0.8}{{\includegraphics{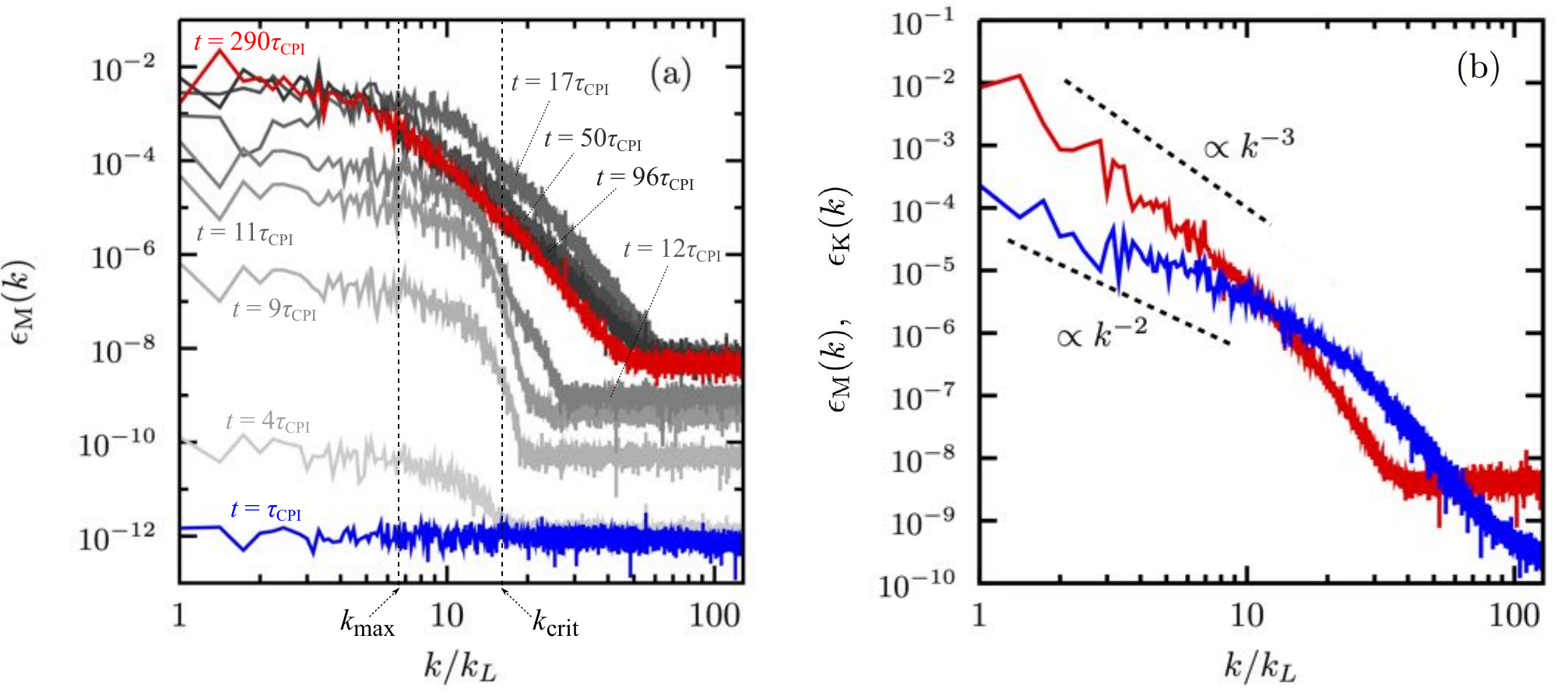}}}  
\caption{(a) Temporal evolution of $\epsilon_{\rm M}(k)$.
  The blue and red curves denote the initial state at $t = \tau_{\rm CPI}$ and nonlinear state at
  $t = 290\tau_{\rm CPI}$. The gray curves are the states between them. (b) $\epsilon_{\rm M} (k)$ (red)
  and $\epsilon_{\rm K} (k)$ (blue) at the saturated state ($t = 500\tau_{\rm CPI}$). The dashed lines
  are reference slopes proportional to $k^{-3}$ and $k^{-2}$. } 
\label{E_M_k}
\end{center}
\end{figure*}
As seen in Fig.~\ref{B}(a), the early evolution of $\langle B^2 \rangle^{1/2}$ agrees with the linear analysis
of the CPI. During this stage, the magnetic field is amplified by a factor of $\mathcal{O}(10^4)$. After the
early exponential growth, it enters the nonlinear stage at $t \simeq 20\tau_{\rm CPI}$. We emphasize that the
saturation amplitude and amplification factor of the magnetic field do {\it not} depend on the strength of the
initial magnetic field but on the initial chirality imbalance (see Sec.~\ref{RS5}). Hence, the magnetic field
can be amplified to the same level as long as we use the same $\xi_B$ as that of the fiducial model, even if a
weaker initial magnetic field is applied. 

The frozen-in property between the plasma and the magnetic field causes the coevolution of the flow field. As
shown in Fig.~\ref{B}(b), $\epsilon_{\rm K}$ rapidly grows when $t \lesssim 100\tau_{\rm CPI}$ and then reaches
a saturation amplitude an order of magnitude smaller than $\epsilon_{\rm M}$. When $t \gtrsim 200 \tau_{\rm CPI}$,
$\epsilon_{\rm K}$ gradually decreases probably due to the viscous dissipation of the small-scale structure of
the flow field. For example, the viscous timescale of the flow structure with the size $l \simeq L/100$ can be
evaluated as $\tau_{\rm vis} = l^2/\nu \simeq 200\tau_{\rm CPI}$. 

Since the chiral MHD turbulence is sustained by the energy conversion from the chirality imbalance to the
magnetic energy, $\langle \xi_B \rangle$ decreases with the increase of the magnetic energy. Shown in
Fig.~\ref{xi_B} is the temporal evolution of $\langle \xi_B \rangle$ normalized by $\xi_{B,{\rm ini}}$. After the
rapid drop stage, it decreases gradually in proportion to $(t/\tau_{\rm CPI})^{-3/5}$ and finally reaches
saturation at $t = t_{\rm sat} \simeq 10^3\tau_{\rm CPI}$ with the floor value
$\xi_{B,{\rm sat}} \equiv \langle \xi_B (t_{\rm sat}) \rangle \simeq 0.077\xi_{B,{\rm ini}}$.

As will be examined in Sec.~\ref{RS5} in detail, the floor value $\xi_{B,{\rm sat}}$ is definitely influenced by
$L$. Using the value of $\xi_{B,{\rm sat}}$ in Table~\ref{setup}, $\lambda_{\rm crit}$ of the CPI at the saturated
stage ($t \simeq 10^3\tau_{\rm CPI}$) is evaluated as
\begin{equation}
\lambda_{\rm crit} = \frac{2\pi}{\xi_{B,{\rm sat}}} \simeq 2.0\times 10^4 = L_0 \;, \label{eq:lambda_crit}
\end{equation}
suggesting that the energy conversion from the chirality imbalance into the magnetic energy is terminated when
$\xi_B $ is reduced to the value at which the unstable wavelength of the CPI becomes comparable to the size of
the calculation domain. 

As expected from the temporal behavior of $\langle \xi_B \rangle$, the spatial structures of ${\bm B}$ and
${\bm v}$ exhibit the inverse energy cascade. Series of snapshots of the distributions of $B_x$ and $v_x$ at
different times on the $x$--$y$ cutting plane at $z = 0$ are shown in Figs.~\ref{B_x} and~\ref{v_x}. The red
and blue tones depict the positive and negative values of the fields. The vertical and horizontal axes are
normalized by $L/2$. While $B_x$ and $v_x$ have small-scale structures in the early evolutionary stage
[panels~(a)--(f)], they evolve as time passes to organize the large-scale structure with the spatial scale
comparable to the size of the calculation domain [panels~(g)--(i)]. It should be stressed that, since there
is no specific direction in our simulation, not only the $x$ component but also the $y$ and $z$ components of
${\bm B}$ and ${\bm v}$ also have similar large-scale structures; see Fig.~\ref{3D} for the 3D structures of
${\bm B}$ and ${\bm v}$, in which their magnitudes at the early and fully nonlinear stages are visualized. 

The inverse-cascade process of the chiral MHD turbulence can be seen in the temporal evolution of the 3D spectrum 
of the magnetic energy density $\epsilon_{\rm M}(k)$ as shown in Fig.~\ref{E_M_k}(a). Here, $\epsilon_{\rm M}(k)$ is
defined by
\begin{equation}
\epsilon_{\rm M}(k) \equiv \frac{1}{2}\sum_{k_x,k_y,k_z}\hat{\mbox{\boldmath $B$}}(\mbox{\boldmath $k$})\cdot\hat{\mbox{\boldmath $B$}}^*(\mbox{\boldmath $k$}) \;, \label{eq:def_em_k}
\end{equation}
where $\hat{\mbox{\boldmath $B$}}(\mbox{\boldmath $k$})$ is the 3D Fourier transform of
$\mbox{\boldmath $B$}(\mbox{\boldmath $x$})$ with $\hat{\mbox{\boldmath $B$}}^*(\mbox{\boldmath $k$})$
being its complex conjugate, and the summation is over all $k_x$, $k_y$ and $k_z$ such that $k_x^2+k_y^2+k_z^2 = k^2$.
The blue and red curves correspond to the initial and nonlinear states (at $t \simeq \tau_{\rm CPI}$ and
$t \simeq 290\tau_{\rm CPI}$). The gray lines are the spectra at the time between these two states. 
In Fig.~\ref{E_M_k}(b), not only $\epsilon_{\rm M}(k)$ (red), but also the spectrum of the kinetic energy density
$\epsilon_{\rm K}(k)$ (blue), which is calculated in a similar manner as Eq.~(\ref{eq:def_em_k}), at the
fully nonlinear stage ($t \simeq 500\tau_{\rm CPI}$) are shown. The horizontal axes are normalized by
$k_L = 2\pi/L$ in both panels. 

$\epsilon_{\rm M}(k)$ begins to grow for $k \lesssim k_{\rm crit}$, which is the linearly unstable regime
of the CPI, and thus it is the energy injection scale for the chiral MHD turbulence. While the magnetic
energy is dominantly contained in the low-wave number modes, it is transferred to the smaller-scale structure
via the direct cascade process. A similar evolution history can also be seen in $\epsilon_{\rm K}(k)$, 
despite a remarkable difference in the spectral slopes, roughly $\epsilon_{\rm M}(k) \propto k^{-3}$ and
$\epsilon_{\rm K}(k) \propto k^{-2}$ in the low-$k$ regime. 
Not only the spatial distribution of the field structures in Figs.~\ref{B_x},~\ref{v_x}, and~\ref{3D} 
but also the spectra in Fig.~\ref{E_M_k} indicate that a more prominent large-scale structure develops in the
magnetic field than in the velocity field. 

\subsection{Dependence on resolution\label{RS3}} 
\begin{figure*}[tbp]
\begin{center}
\scalebox{0.75}{{\includegraphics{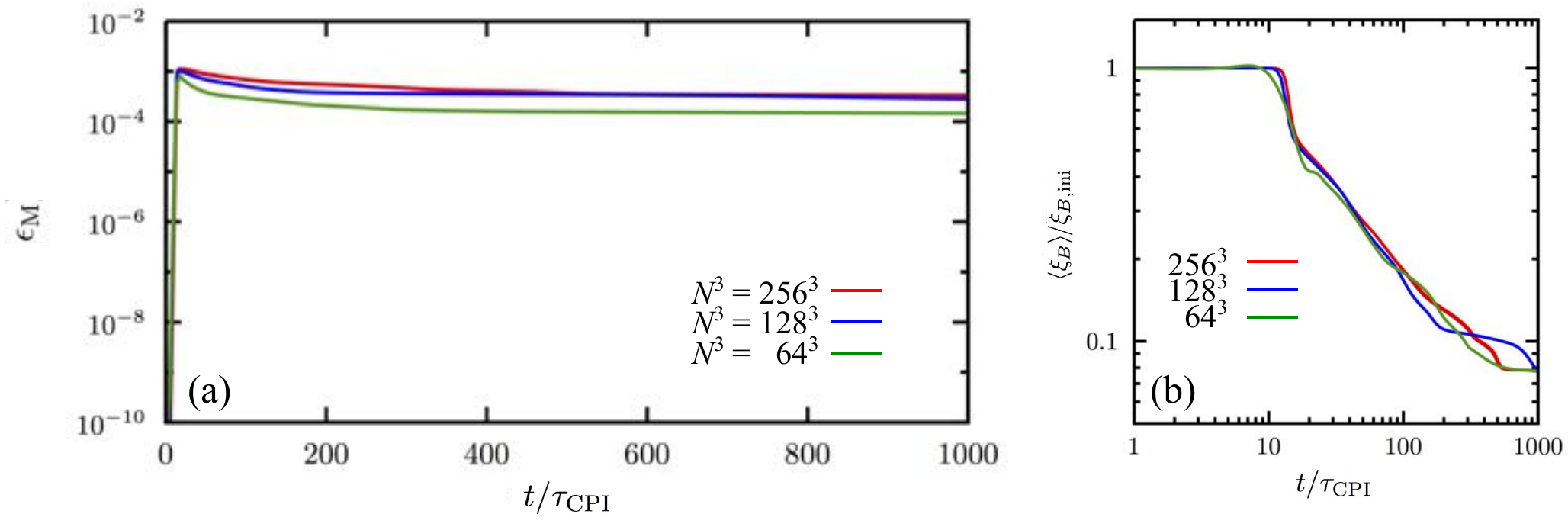}}}  
\caption{Temporal evolutions of (a) $\epsilon_{\rm M}$ and (b) $\langle \xi_B \rangle $ for the models
  with different numerical resolutions. The red, blue, and green lines are for the models with $N = 256^3$,
  $128^3$, and $64^3$, respectively.} 
\label{xi_B-N}
\end{center}
\end{figure*}
\begin{figure*}[tbp]
\begin{center}
\scalebox{0.65}{{\includegraphics{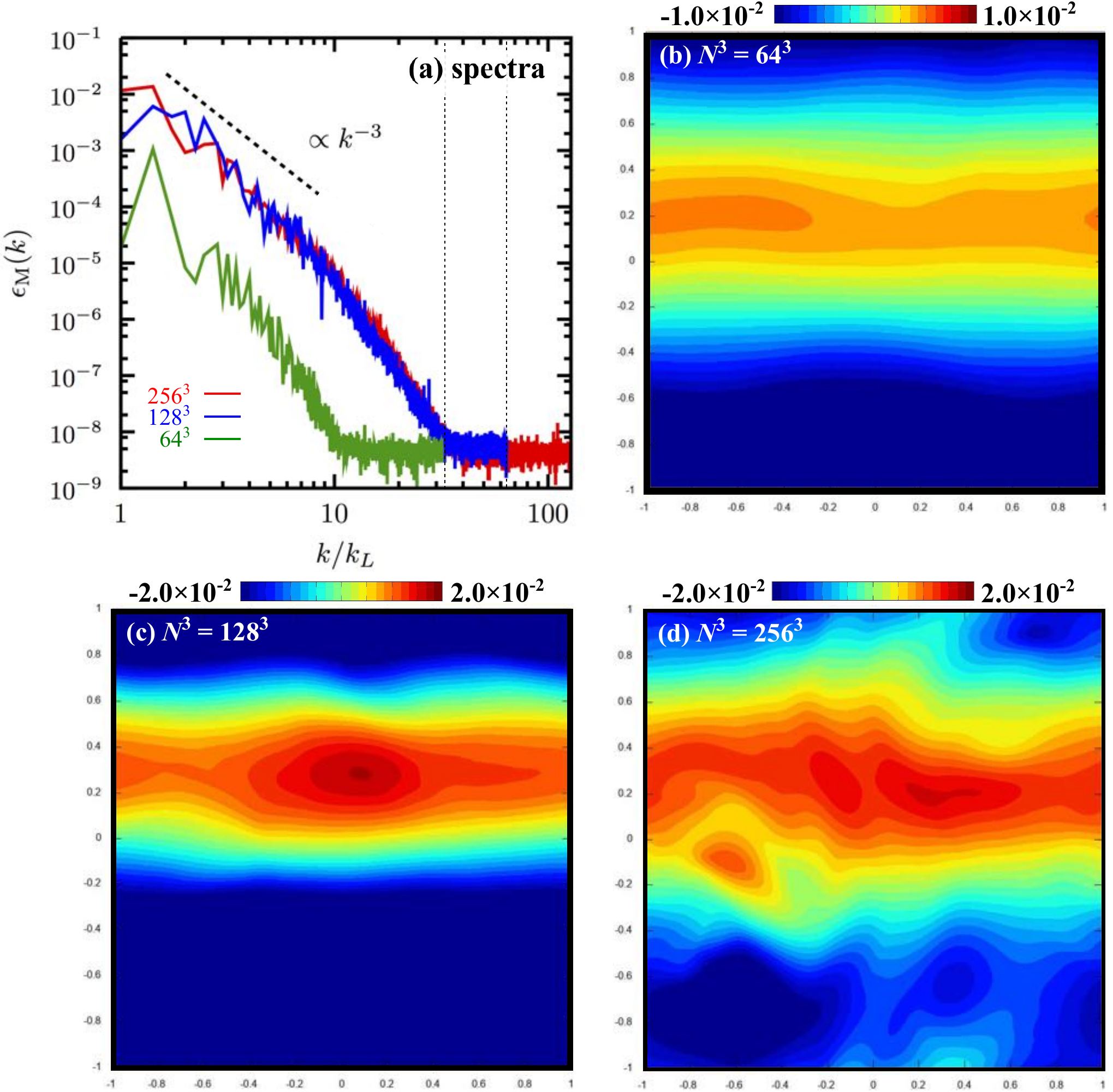}}}  
\caption{(a) $\epsilon_{\rm M}(k)$ for the models with different resolutions. The dashed line is
  a reference slope $\propto k^{-3}$. The $x$--$y$ distributions of $B_x$ at the saturated stage
  for the models with (b) $N = 64^3$ (model 3), (c) $128^3$ (model 2), and (d) $256^3$ (model 1).} 
\label{E_M_k-N}
\end{center}
\end{figure*}
To conduct the parametric study, we need to know how many grids are required at least to correctly capture the
behavior of the chiral MHD turbulence. The convergence is checked by comparing the models with different
resolutions. Models~2 and~3 have the number of grids $N^3 = 128^3$ and $N^3 = 64^3$ with keeping the other
parameters the same as in the fiducial model with $N^3 = 256^3$ (model~1). 

Figure~\ref{xi_B-N} shows the temporal evolution of (a) $\epsilon_{\rm M}$ and (b) $\langle \xi_B \rangle $ for
the models with different resolutions. The red, blue, and green lines denote models~1--3, respectively, in
both panels. We find that the saturation amplitude of $\epsilon_{\rm M}$ converges when $N \gtrsim 128$. 
Model~3 has an insufficient resolution, yielding the lower saturation amplitude. In contrast, $\xi_{B,{\rm sat}}$ 
is not affected by the resolution, suggesting again that it is determined numerically by the size of the
calculation domain. 

The convergence of the numerical result can be confirmed more quantitatively by comparing the spectra of the
models. Shown in Fig.~\ref{E_M_k-N}(a) is $\epsilon_{\rm M}(k)$ at the saturated stage for the models with
different resolutions. The red, blue, and green lines correspond to models~1--3, respectively. 
Models~1 and~2 with $N^3 = 256^3$ and $128^3$ have almost the same spectral profiles and amplitudes. However,
the spectrum of the model~3 with $N^3 = 64^3$ deviates significantly from them, verifying that it is insufficient
for correctly capturing the chiral MHD turbulence. 

The distributions of $B_x$ at the saturated stage on the $x$--$y$ cutting plane at $z = 0$ for the models 1--3
are presented in Figs.~\ref{E_M_k-N}(b)--(d). The red and blue tones depict the positive and negative values of
the magnetic field. While the large-scale magnetic structure with the wavelength comparable to the box size is
a common outcome of the nonlinear evolution of the chiral MHD turbulence as a result of the inverse cascade,
the strength of the magnetic field is an order of magnitude weaker in the lowest resolution model (model~3)
than in the sufficiently resolved models (models~1 and 2).

From these results, the convergence seems to be achieved when 
\begin{equation}
\lambda_{\rm CPI}/\Delta \gtrsim 7 \;. \label{convergence}
\end{equation}
At the same time, our results imply that even lower resolution calculation is acceptable as long as 
we are only interested in $\xi_{B,{\rm sat}}$.

\subsection{Dependence on resistivity \label{RS4}} 
\begin{figure}[htbp]
\begin{center}
\scalebox{0.42}{{\includegraphics{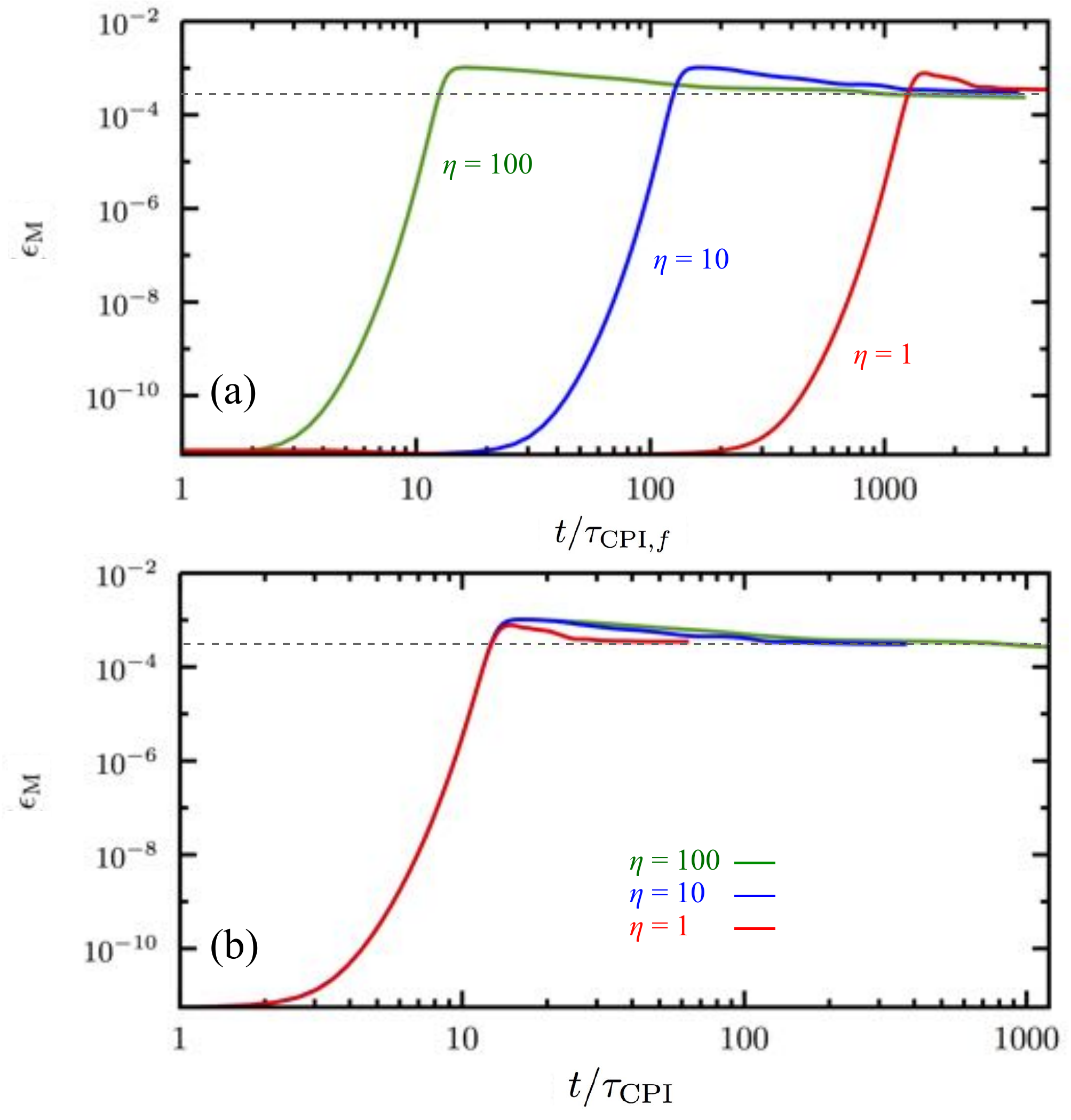}}}  
\caption{Temporal evolutions of $\epsilon_{\rm M}$ for the models with different values of $\eta$.
  The red, blue, and green lines correspond to the models with $\eta = 1$, $10$, and $100$, respectively.
  The normalization of the horizontal axis is different between panels (a) and (b). In panel (a),
  the horizontal axis is normalized by $\tau_{\rm CPI,f}$. In panel (b), the normalization unit is
  $\tau_{\rm CPI}$.} 
\label{E_M-eta}
\end{center}
\end{figure}
One of the key parameters for the CPI and its driven MHD turbulence is the resistivity $\eta$. As described
in Fig.~\ref{CPI}(a), one of the interesting features of the CPI is that its linear growth rate becomes higher
with the increase of $\eta$, while it suppresses the MHD turbulence in most cases of the conventional nonchiral
MHD. The effects of $\eta$ on the nonlinear behavior of the chiral MHD turbulence is of our interest here. We
run the models~4 and~5 with $\eta = 10$ and $1$ and then compare them with the model~2 with $\eta = 100$. We
take the resolution and the other parameters to be the same.

In Fig.~\ref{E_M-eta}, we show the temporal evolution of $\epsilon_{\rm M}$ for the models with $\eta = 1$ (red),
$10$ (blue) and $100$ (green), respectively. The difference between panels (a) and (b) is the normalization unit
of time. In panel~(a), the simulation time of each model is normalized in common by $\tau_{\rm CPI}$ for the
model with $\eta = \eta_0$, $\tau_{{\rm CPI},f} \equiv 4(\eta_0\xi_{B,{\rm ini}}^2)^{-1}$. In contrast, in panel~(b),
it is normalized by $\tau_{\rm CPI}$ evaluated with $\eta$ of each model.

\begin{figure}[htbp]
\begin{center}
\scalebox{0.42}{{\includegraphics{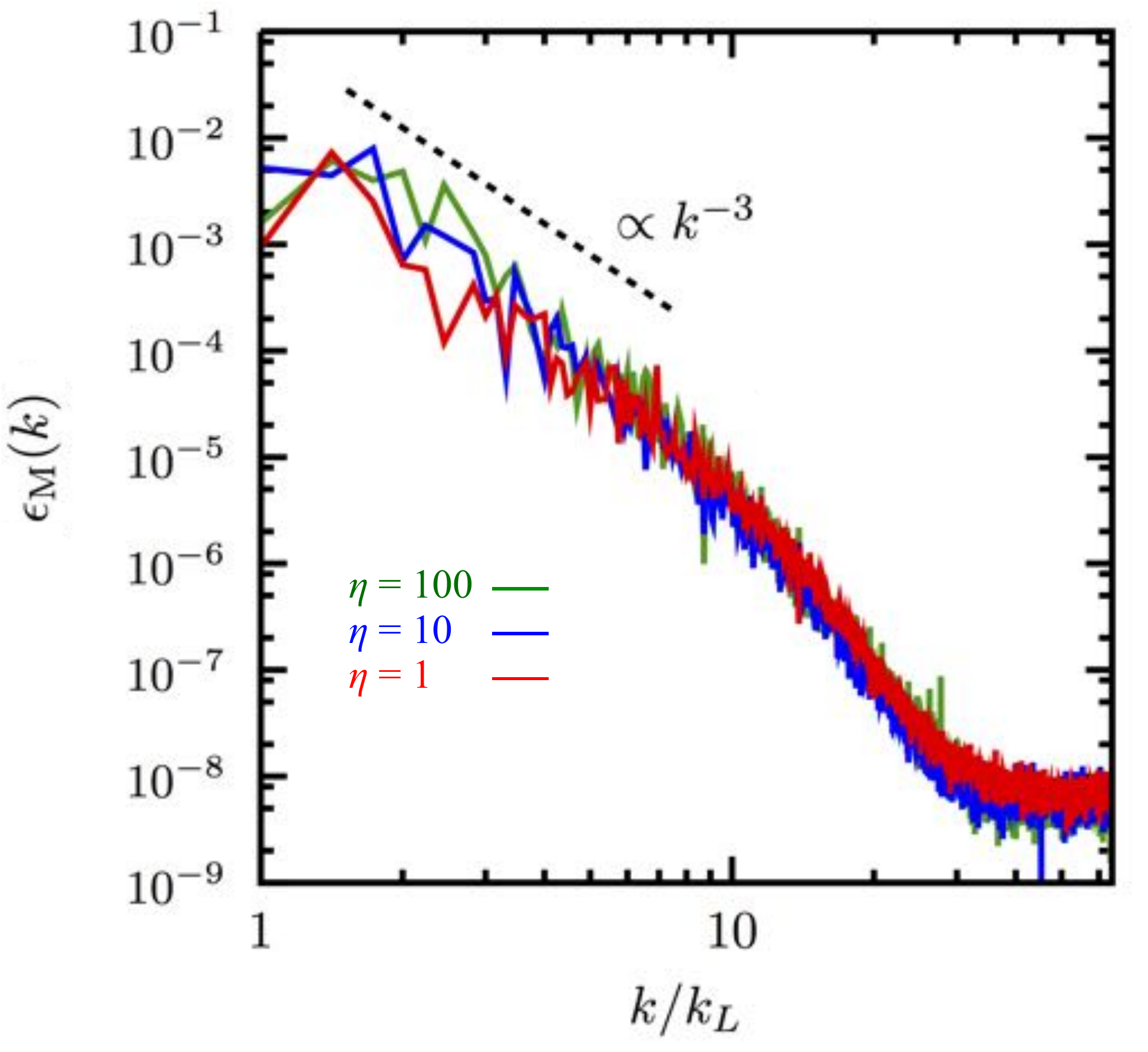}}}  
\caption{$\epsilon_{\rm M} (k)$ at the saturated stage for the models with different values of $\eta$.
  The red, blue, and green lines correspond to the models with $\eta = 1$, $10$, and $100$, respectively.
  The dashed line is a reference slope proportional to $k^{-3}$.} 
\label{E_M_k-eta}
\end{center}
\end{figure}
\begin{figure}[htbp]
\begin{center}
\scalebox{0.42}{{\includegraphics{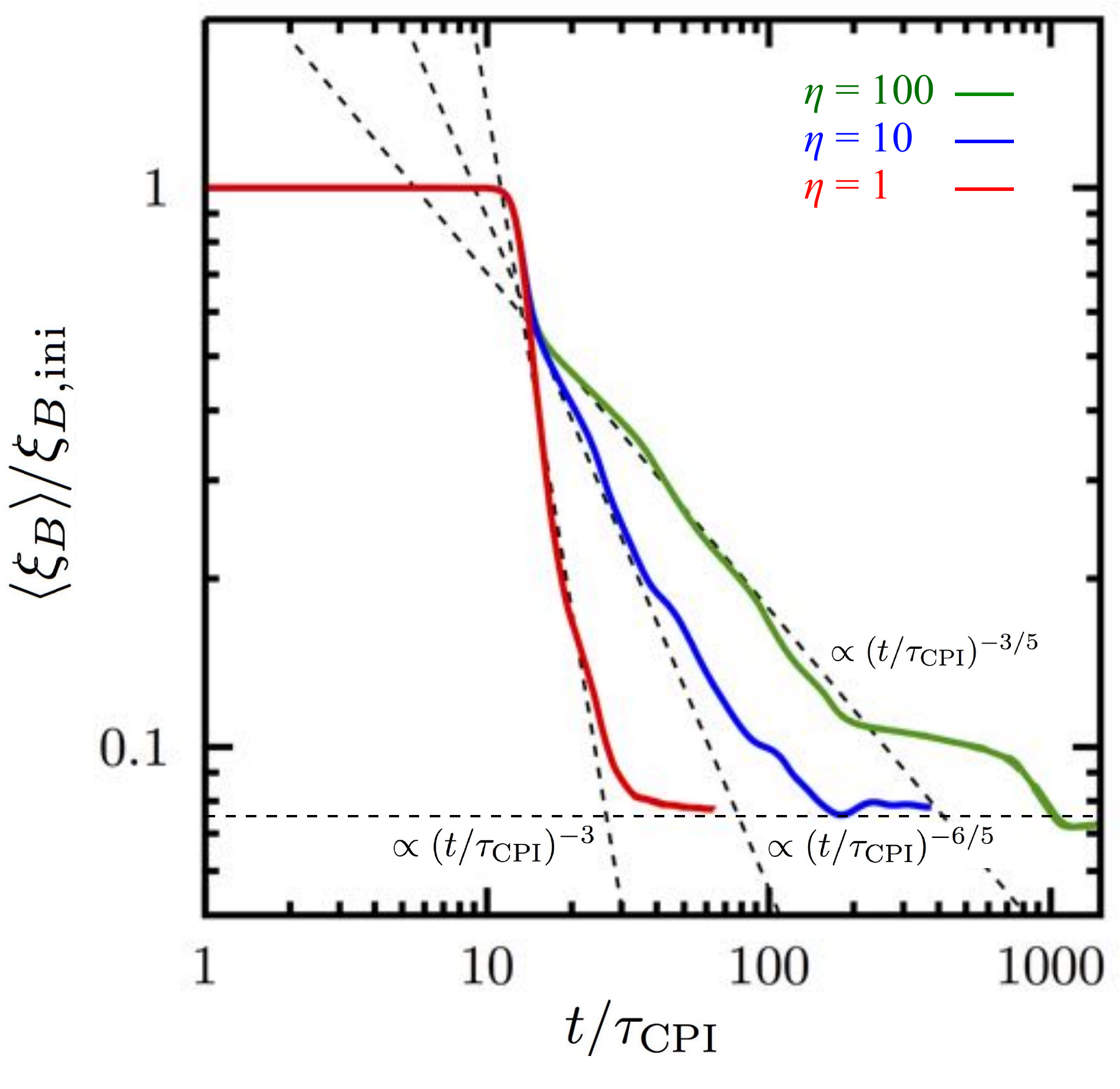}}}  
\caption{Temporal evolutions of $\langle \xi_B \rangle$ normalized by $\xi_{B,{\rm ini}}$ for the models with
  different values of $\eta$. The red, blue, and green lines correspond to the models with $\eta = 1$, $10$,
  and $100$, respectively.} 
\label{xi_B-eta}
\end{center}
\end{figure}

As seen in panel (a), the actual growth time of the chiral MHD turbulence becomes shorter with increasing
$\eta$, which is consistent with the linear analysis of the CPI [see Eq.~(\ref{eq:CPI})]. However, when
normalizing the simulation time by $\tau_{\rm CPI}$ of each model, the evolution history until the nonlinear
stage is identical between the models. In addition, the saturation amplitude of $\epsilon_{\rm M}$ is also roughly
the same between the models in spite of the difference of $\eta$. 

The independence of the nonlinear behavior of the chiral MHD turbulence on $\eta$ can be seen even in the
comparison of the spectra. Shown in Fig.~\ref{E_M_k-eta} is $\epsilon_{\rm M}(k)$ at the saturated stages for
these models. The red, blue, and green lines denote the models with $\eta = 1$, $10$ and $100$, respectively.
Regardless of the size of $\eta$, the chiral MHD turbulence exhibits a similar spectral property. All the results
suggest that $\eta$ does not have a strong impact on the nonlinear behavior of the chiral MHD turbulence though 
it changes the linear growth rate of the CPI. 

The evolution history of $\langle \xi_B \rangle $ might be one of the few differences between these models. The
temporal evolutions of $\xi_B$ for these models are shown in Fig.~\ref{xi_B-eta}. Three dashed lines are the
reference slopes proportional to $(t/\tau_{\rm CPI})^{-3/5}$, $(t/\tau_{\rm CPI})^{-6/5}$, and $(t/\tau_{\rm CPI})^{-3}$,
respectively. While $\xi_{B,{\rm sat}}$ is almost the same, $t_{\rm sat}$ is difference between them. With decreasing
$\eta$, the normalized time required for the saturation becomes shorter. This might be because the higher magnetic
diffusion makes the magnetic structure harder to grow at the nonlinear stage. 

\subsection{Dependence on box size \label{RS5}} 
As discussed in Sec.~\ref{RS2}, $\xi_{B,{\rm sat}}$, which is responsible for the conversion efficiency of the
chirality imbalance into the magnetic energy, is expected to be determined by the size of the calculation domain
in our local-box model. To verify this, we examine the response of the chiral MHD turbulence to the change of $L$.
The models with $L = 5L_0,\ L_0/2$, and $L_0/5$ (models 6, 7, and 8) are compared with model 5 with $L = L_0$.
We keep, as far as possible, the ratio $\lambda_{\rm CPI}/\Delta$ constant rather than the number of grids, except
for the largest box model with $L = 5L_0$ (model~6) in which case the higher resolution of $N^3 = 640^3$ is required. 
As was shown in Sec.~\ref{RS3}, the insufficient resolution does not matter as long as we focus on $\xi_{B,{\rm sat}}$. 
The other physical parameters are kept unchanged from model~5 with the fiducial box size.

\begin{figure}[htbp]
\begin{center}
\scalebox{0.42}{{\includegraphics{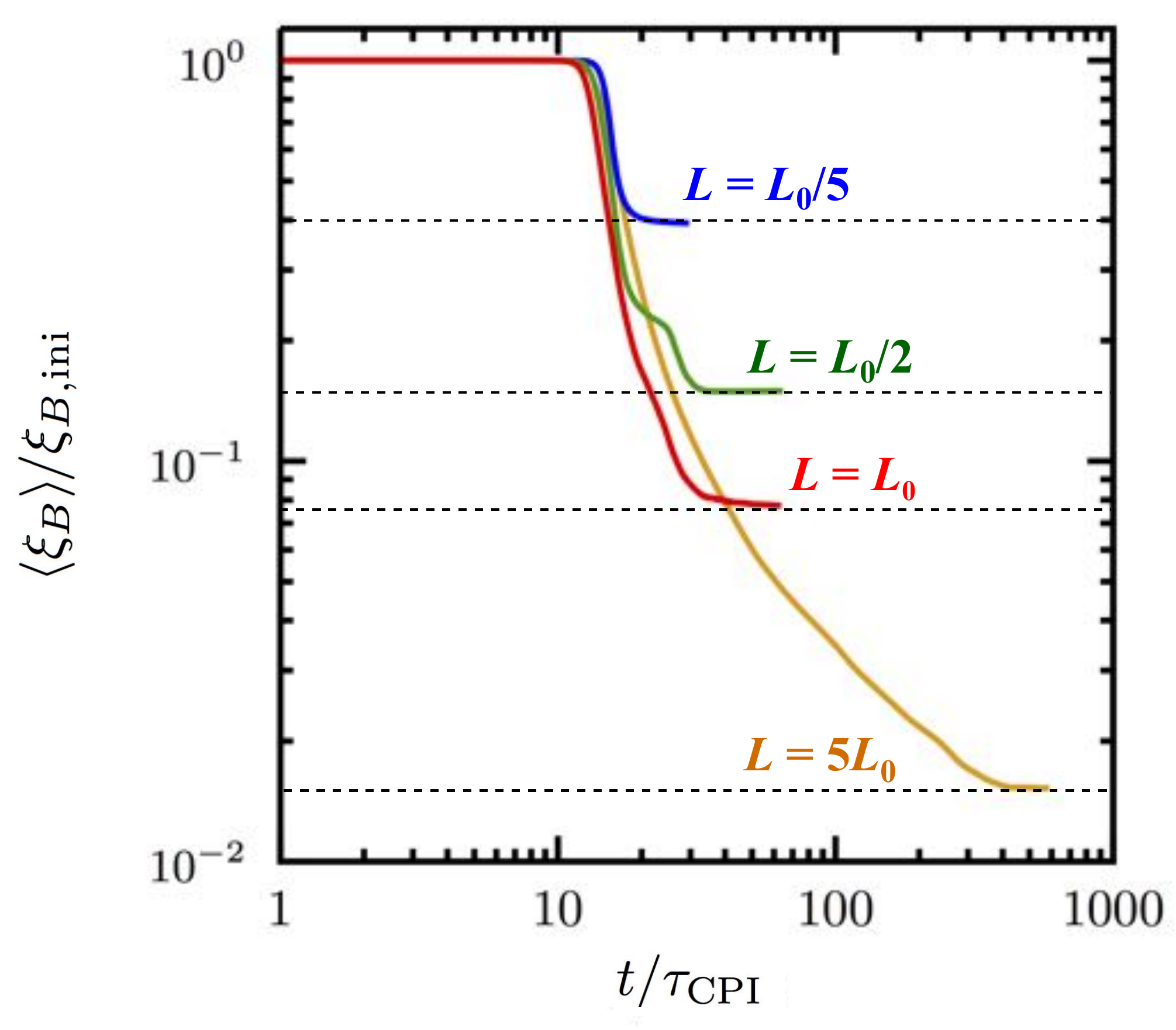}}}  
\caption{Temporal evolutions of $\langle \xi_B \rangle$ for the models with the different $L$.
  The blue, green, red, and orange lines denote the models with $L=5L_0$, $L_0$, $L_0/2$, and $L_0/5$. The
  vertical axis is normalized by the initial value of $\langle \xi_B \rangle$ for each model.}
\label{xi_B-L}
\end{center}
\end{figure}

\begin{figure*}[htbp]
\begin{center}
\scalebox{0.65}{{\includegraphics{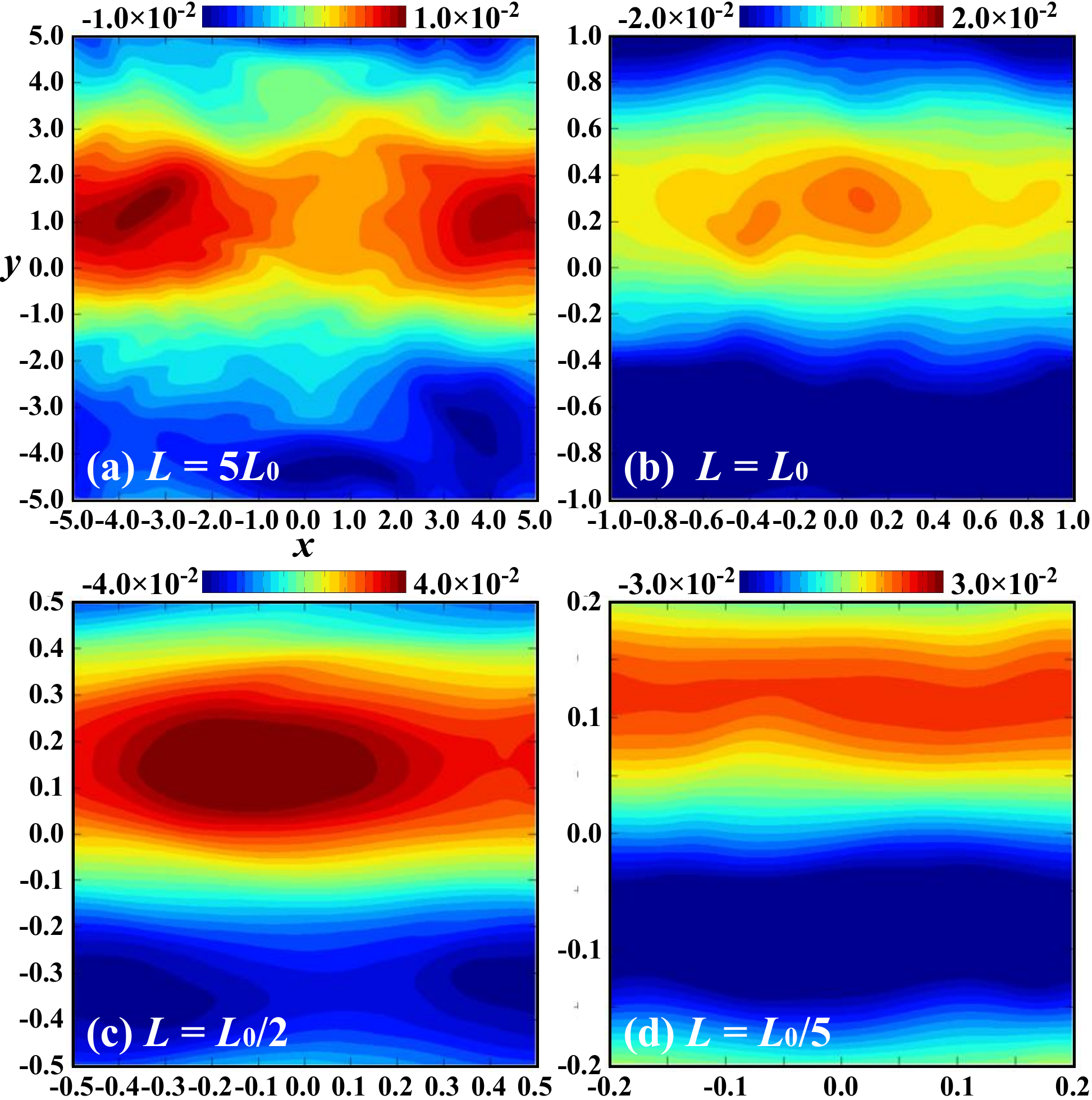}}}  
\caption{Distributions of $B_x$ at the saturated stages on the $x$--$y$ cutting plane at $z=0$ for the models
  with (a) $L = 5L_0$, (b) $L = L_0$, (c) $L = L_0/2$, and (d) $L = L_0/5$. The axes are normalized by $L_0/2$
  for all the models.} 
\label{B_x-L}
\end{center}
\end{figure*}

Shown in Fig.~\ref{xi_B-L} is the temporal evolution of $\langle \xi_B \rangle$ for each model until the saturation.
The orange, red, green, and blue lines are for the models with $L = 5L_0$, $L_0$, $L_0/2$, and $L_0/5$, respectively.
The distributions of $B_x$ on the $x$--$y$ cutting plane at $z = 0$ at the saturated stage are also demonstrated for
each model in Fig.~\ref{B_x-L}. The axes of all the panels are normalized by $L_0/2$ of the fiducial model.

We find that $\xi_{B,{\rm sat}}$ is different when $L$ is varied, despite the same physical parameters except for $L$. 
It is inversely correlated with $L$, i.e., $\xi_{B,{\rm sat}} \propto L^{-1}$ (see Table~\ref{setup}), and provides the
critical wavelength of the CPI comparable to the box size of each model,
$\lambda_{\rm crit} = 2\pi/\xi_{B,{\rm sat}} \simeq L$. In Fig.~\ref{B_x-L}, we can indeed confirm that the spatial
structure of the magnetic field is comparable to the domain size. All these results verify our hypothesis that
$\xi_{B,{\rm sat}}$ is restricted numerically by the calculation domain, and furthermore implies that the structures
of ${\bm B}$ and ${\bm v}$ can evolve to the macroscopic scale comparable to the size of the PNS if we can enlarge
the simulation domain to the system scale with keeping the sufficient resolution.


\subsection{Dependence on axial charge density \label{RS6}} 

\begin{figure*}[htbp]
\begin{center}
\scalebox{0.85}{{\includegraphics{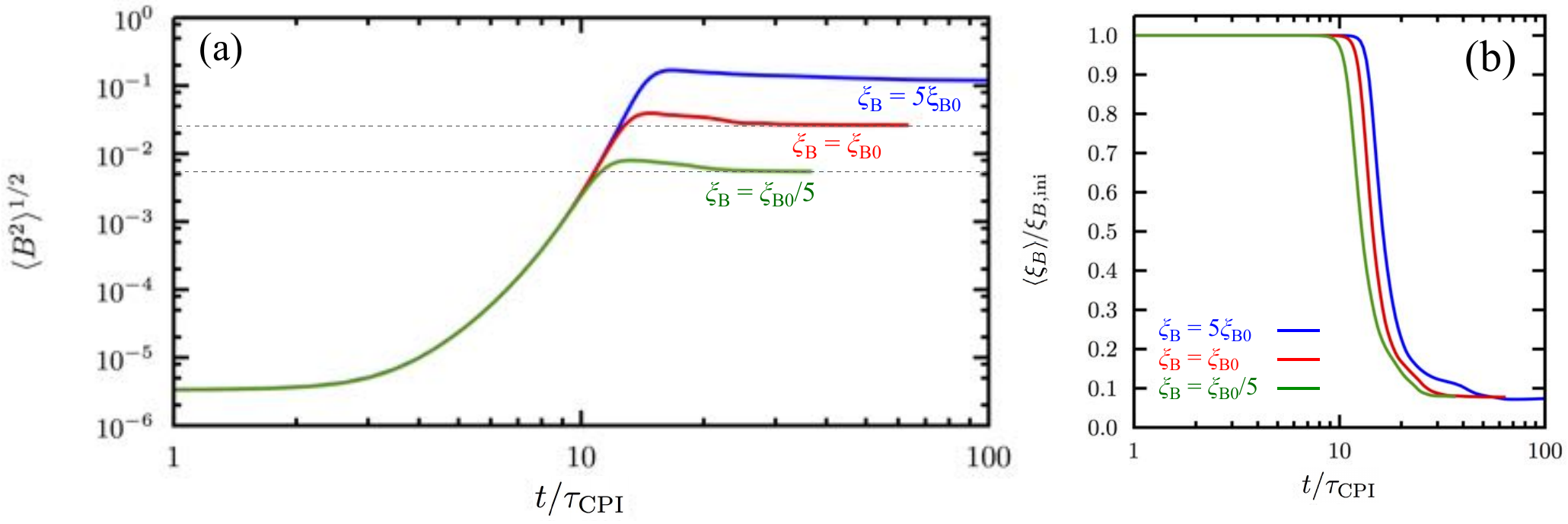}}}  
\caption{Temporal evolutions of (a) $\langle B^2\rangle^{1/2}$ and (b) $\xi_B$ for the models with the different
  initial values of $n_{\rm A}$. The red, blue and green lines denote the models with
  $\xi_{B,{\rm ini}} = 5\xi_{B0}$, $\xi_{B0}$, and $\xi_{B0}/5$.} 
\label{B-n_A}
\end{center}
\end{figure*}
\begin{figure}[htbp]
\begin{center}
\scalebox{0.4}{{\includegraphics{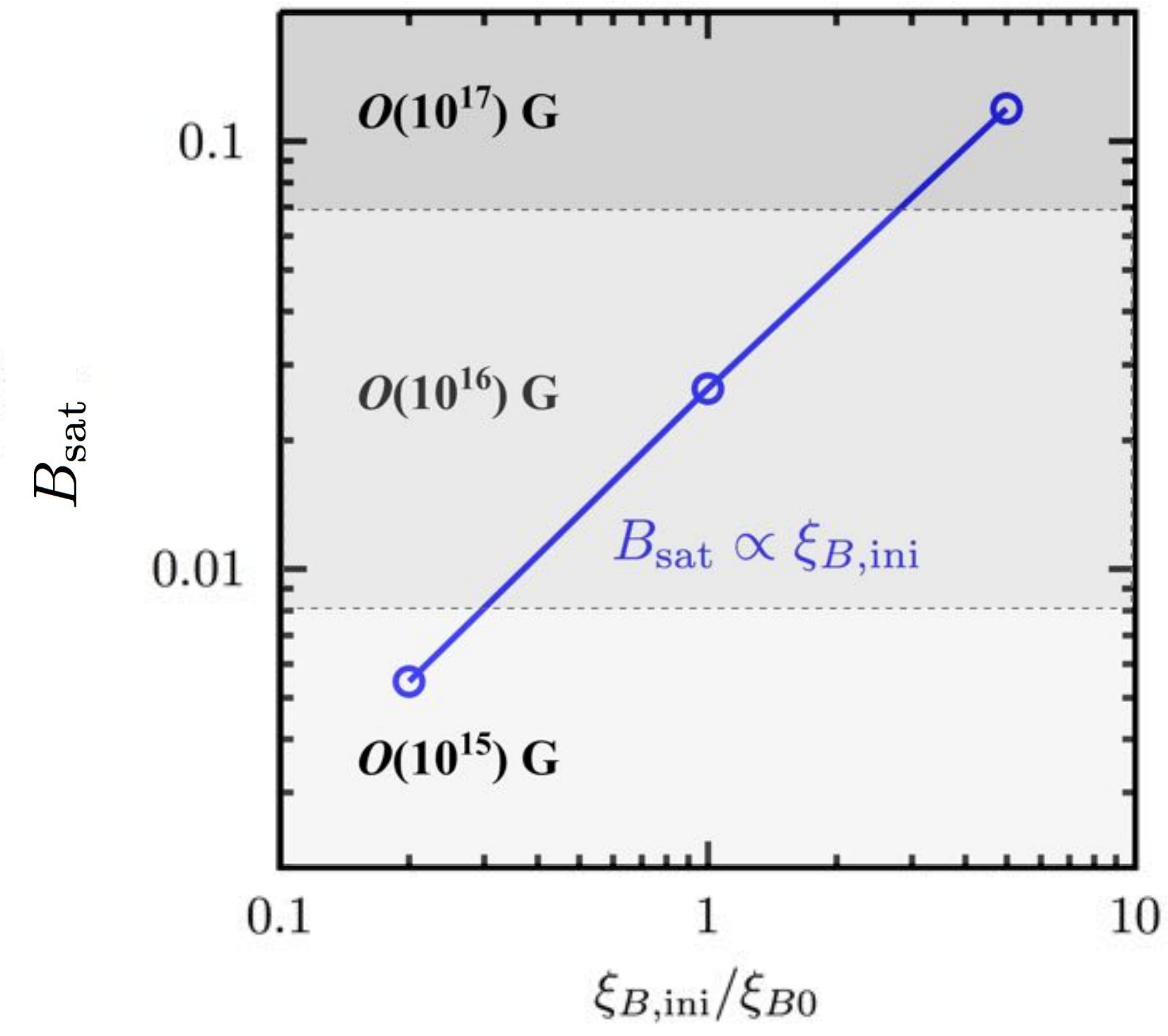}}}  
\caption{Scaling relation between $B_{\rm sat}$ and $\xi_{B,{\rm ini}}$ normalized by the fiducial value $\xi_{B0}$.} 
\label{B-xi_B}
\end{center}
\end{figure}

Finally, we study the dependence of the behavior of the chiral MHD turbulence on the initial value of $n_{\rm A}$. 
Remember that $n_{\rm A}$ is directly related to $\xi_B$ [see Eq.~(\ref{eq:xi_B})], and thus, it is the most
important parameter in our simulation. In models 9 and 10, we set $n_{\rm A} = 0.416$ and $0.020$,
i.e., $\xi_{B} = 5\xi_{B0}$ and $\xi_{B0}/5$, respectively. For a fair comparison between the models, we need to
keep the ratio $L/\lambda_{\rm crit}$ constant because, as discussed in Sec.~\ref{RS5}, $L/\lambda_{\rm crit}$ affects
$\xi_{B,{\rm sat}}$ or the energy conversion efficiency. Therefore, the box sizes $L = L_0/5$ and $L = 5L_0$ are
adopted for models 9 and 10, correspondingly. The other parameters are kept unchanged from model~5 with
$\xi_B = \xi_{B0}$ and $L = L_0$. 

Shown in Figs.~\ref{B-n_A}(a) and (b) are the temporal evolutions of $\langle B^2 \rangle^{1/2}$ and
$\langle \xi_B \rangle$ for these models. The blue, red, and green lines denote the models with
$\xi_B = 5\xi_{B0}$, $\xi_{B0}$ and $\xi_{B0}/5$, respectively. Note that, in both panels, the simulation time
is normalized by $\tau_{\rm CPI}$ of each model. The vertical axis of the panel (b) is normalized by
$\xi_{B,{\rm ini}}$ of each model. 

At the early stage $t \lesssim 10\tau_{\rm CPI}$, the evolution history of $\langle B^2 \rangle^{1/2}$ is consistent
with the linear analysis of the CPI and does not depend on $\xi_{B,{\rm ini}}$. However, at the nonlinear stage,
there exists a remarkable difference despite $\langle \xi_B \rangle /\xi_{B,{\rm ini}}$ being almost the same
[see panel(b)]. We can evaluate $B_{\rm sat} \equiv \langle \mbox{$B$}^2(t_{\rm sat}) \rangle^{1/2}$ from the time
average of $\langle B^2 \rangle^{1/2}$ at the nonlinear stage, as plotted in Fig.~\ref{B-xi_B} as a function of
$\xi_{B,{\rm ini}}$. From this, we find the scaling relation, 
\begin{eqnarray}
B_{\rm sat} \propto \xi_{B,{\rm ini}}\;, \label{scaling}
\end{eqnarray}
indicating that the magnetic field strength maintained by the chiral MHD turbulence is a linear function of the
total amount of $\xi_{B,{\rm ini}}$ generated in the supernova core. 

\section{Discussion}
\label{sec:discussion}
\subsection{Turbulence scaling}
The spectrum of the chiral MHD turbulence was previously discussed for high-temperature plasmas in the early
Universe in Ref.~\cite{Brandenburg:2017rcb}. They observed the weak turbulence scaling with
$\epsilon_{\rm M}(k) \propto k^{-2}$ in the turbulent scales $k \lesssim \xi_B $ in the magnetically dominated
turbulence, where the flow field is a consequence of driving by Lorentz force \cite{Galtier:2000ce,Brandenburg:2014mwa}. 
Note that our spectral scaling $\epsilon_{\rm M}(k) \propto k^{-3}$ and $\epsilon_{\rm K}(k) \sim k^{-2}$
in the low-$k$ regime in Fig.~\ref{E_M_k}(b) at the fully nonlinear stage is different from theirs.

The reason for this difference can be explained by the evolution of $\langle \xi_B \rangle$. 
While the spectrum in Ref.~\cite{Brandenburg:2017rcb} is the case {\it before} $\langle \xi_B \rangle$ reaches
the floor value, our spectra in Fig.~\ref{E_M_k}(b) are derived {\it after} that, where $\lambda_{\rm crit} \sim L$,
and the turbulent scales are absent. Since the chiral MHD turbulence has smaller scales than the instability scale,
its spectrum shows the steeper slope than in the turbulent scales.

\begin{figure}[htbp]
\begin{center}
\scalebox{0.4}{{\includegraphics{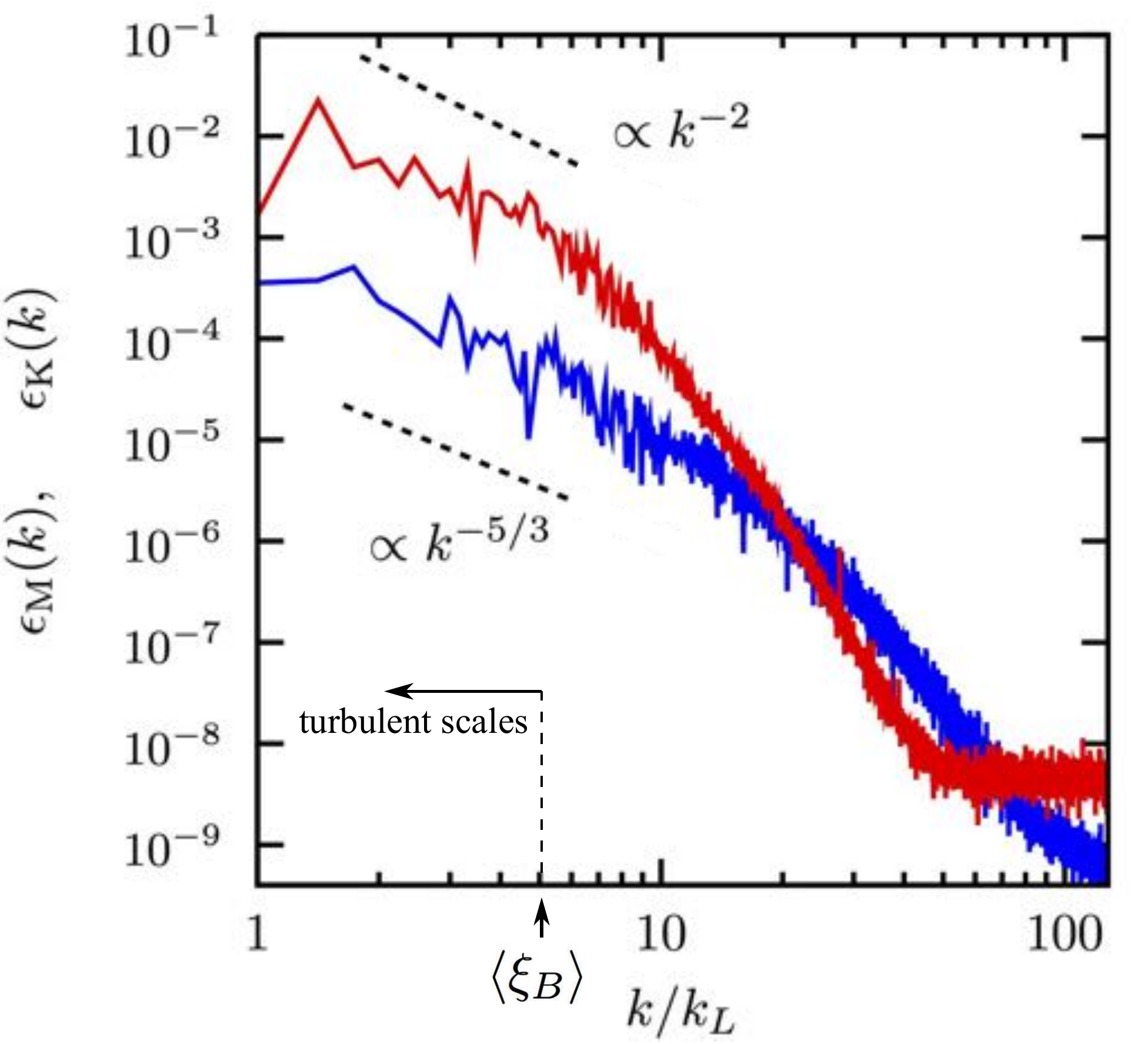}}}  
\caption{$\epsilon_{\rm M} (k)$ (red) and $\epsilon_{\rm K} (k)$ (blue) at $t = 300\tau_{\rm CPI}$ before
  $\epsilon_{\rm K}$ begins to decrease for the fiducial model. The arrow shows $\langle \xi_B \rangle$
  at the corresponding time. The dashed lines are reference slopes proportional to $k^{-2}$ and $k^{-5/3}$.} 
\label{E_M_E_K_before}
\end{center}
\end{figure}

Figure~\ref{E_M_E_K_before} shows the spectra of $\epsilon_{\rm M}(k)$ and $\epsilon_{\rm K}(k)$ before
$\langle \xi_B \rangle$ reaches the floor value ($t \simeq 300\tau_{\rm CPI}$) for the fiducial model.
It can be seen that $\epsilon_{\rm M}(k)$ is roughly proportional to $k^{-2}$ in the range
$k \lesssim \langle \xi_B \rangle$, as is consistent with the weak turbulence scaling discussed
in Ref.~\cite{Brandenburg:2017rcb}. In addition, we also observe that $\epsilon_{\rm K}(k) \sim k^{-5/3}$ in this
stage. This suggests that the chiral MHD turbulence has weak turbulence scalings, $\epsilon_{\rm M}(k) \propto k^{-2}$
and $\epsilon_{\rm K}(k) \propto k^{-5/3}$, in the regime $k \lesssim \langle \xi_B \rangle$ for high-density matter
in the supernova core as well.

\subsection{Chiral vortical effect}
\label{sec:CVE}
In this study, we have ignored the CVE just for simplicity. 
Including the CVE modifies the induction equation as \cite{Yamamoto:2015gzz}
\begin{equation}
\partial_t \mbox{\boldmath $B$} = \bm \nabla \times (\mbox{\boldmath $v$} \times \mbox{\boldmath $B$})
  + \eta \bm \nabla^2 \mbox{\boldmath $B$}  + \eta \bm \nabla \times \left( \xi_{B}\mbox{\boldmath $B$} + \xi_\omega \mbox{\boldmath $\omega$} \right)  \;. \label{eq:B_CVE}
\end{equation}
When taking account of the CVE, the energy reconversion from the flow field to the magnetic field is naively expected, 
because a helical flow field is generated as a consequence of the chiral transport phenomena \cite{Yamamoto:2015gzz}. 
The ratio $\epsilon_{\rm M}/\epsilon_{\rm K}$ may be varied depending on $\xi_\omega$.

In the actual PNS system, the CVE may provide significant change for the chiral MHD turbulence, since the helical
and vortical flow motions should be excited not only locally but also globally through several macroscopic effects
mainly due to the global rotation and the stratified structure; 
see, e.g., Refs.~\cite{Brandenburg:2004jv,masada+14b,masada+16,masada+13}.
In the region with the magnetic field parallel to the vortical axis, the CVE should enhance the magnetic energy by
the conversion from the fluid helicity to the magnetic helicity. In an opposite way, it should be possible that the
magnetic energy is reduced by the CVE in the region with the magnetic field antiparallel to the vortical axis.
A quantitative understanding of the CVE in the actual global system of the PNS is beyond the scope of this paper
and is a target of our future work.

\section{Summary}
\label{sec:conclusion}
In this paper, we have performed 3D numerical simulations of the chiral MHD turbulence, driven by the CPI,
in the vicinity of the PNS in the supernova core. We adopted a local Cartesian model which zooms in on a small
patch of the PNS. Our findings are summarized as follows.

\begin{enumerate}
\item{The magnetic field is amplified exponentially in accordance with the linear analysis of the CPI in the early
  evolutionary stage and then enters the nonlinear stage at around $t \simeq \mathcal{O}(10)\tau_{\rm CPI}$. Not only
  the magnetic field, but the flow field also coevolves due to the frozen-in property of the plasma and magnetic
  field. The kinetic energy of the chiral MHD turbulence is an order of magnitude smaller than the magnetic one at
  the nonlinear stage.}

\item{The magnetic and velocity fields exhibit the inverse energy cascade. While they have small-scale structures
  in the early evolutionary stage, they evolve to organize the large-scale structure with the spatial scale
  comparable to the size of the calculation domain. This conforms with what the linear analysis of the CPI predicts,
  i.e., the typical wavelength of the CPI is proportional to the chiral magnetic conductivity,
  $\lambda_{\rm crit} \propto \xi_B$, and it becomes longer as $\xi_B$ decreases with time. At the saturated stage,
  the Fourier spectra of the magnetic and kinetic energy densities have slopes in proportion to $k^{-3}$ and $k^{-2}$,
  respectively.}

\item{Two numerical parameters impact the chiral MHD turbulence: one is the resolution and the other is the box size.
  The sufficient condition for resolving the chiral MHD turbulence is $\lambda_{\rm crit}/\Delta \gtrsim 7$, where
  $\Delta$ is the grid size. While the lower resolution run provides the lower amplitude of the chiral MHD turbulence,
  the floor value of $\xi_B$ is predominantly determined by the size of the calculation domain (less sensitive to the
  numerical resolution). The larger the size of the calculation domain, the lower the floor value of $\xi_B$ is.
  Therefore, the spatial structures of the magnetic and flow fields become larger with increasing the box size,
  implying that they can evolve to a macroscopic scale comparable to the size of the PNS if the calculation domain
  is enlarged to the system scale with keeping the sufficient resolution.}

\item{One of the key physical parameters for the CPI is the resistivity $\eta$ because the growth rate of the CPI
  becomes larger with increasing $\eta$. We found from the parametric study that the size of $\eta$ does not have a
  significant impact on the chiral MHD turbulence itself, though the time required for the saturation of the CPI
  largely depends on it.}

\item{The strength of the chiral MHD turbulence is essentially determined by the initial axial charge density. The
  scaling relation between the saturated value of the mean magnetic-field strength and the initial value of the chiral
  magnetic conductivity is given by $B_{\rm sat} \propto \xi_{B,{\rm ini}}$. This indicates that the magnetic-field
  strength maintained by the chiral MHD turbulence is a linear function of the total amount of $\xi_B$ generated in
  the supernova core.}
\end{enumerate}

Our results suggest that the chiral effects of leptons would impact the dynamics of the PNS formation and supernova
explosions by driving the strong MHD turbulence with the strong magnetic field. The next step of our study is modeling
and taking account of the chiral effects into the global multidimensional MHD supernova simulations 
(e.g., Refs.~\cite{Mosta:2014jaa,Obergaulinger:2017qno}) to elucidate their dynamical impacts more quantitatively. 
In particular, it would be important to incorporate the contributions of the chiral transport of neutrinos
\cite{Yamamoto:2015gzz}, since they carry away most of the gravitational energy of an original massive star.

It should be remarked here that neutrinos are not always in thermal equilibrium, especially outside the supernova
core, where hydrodynamics for neutrinos is not applicable. To take into account the effects of the
left-handed-ness of neutrinos away from thermal equilibrium, one needs to use the so-called chiral kinetic theory
\cite{Son:2012wh,Stephanov:2012ki,Chen:2012ca}, instead of the conventional kinetic theory (Boltzmann equation).
Such a direction is deferred to future work.

\section*{Acknowledgement}
The authors thank an anonymous referee for constructive comments on the manuscript.
This work was supported by JSPS KAKENHI Grants No.~JP15K17611, No.~JP18H01212, No.~JP18K03700, No.~JP15KK0173, No.~JP17H06357, 
No.~JP17H06364, No.~JP17H01130, No.~JP17K14306, No.~JP17H05206, and No.~JP16K17703. K.K. acknowledges support from the Central Research
Institute of Fukuoka University (Grants No.~171042, and No.~177103). N.Y. was supported by MEXT-Supported Program for the Strategic
Research Foundation at Private Universities, ``Topological Science" (Grant No.~S1511006). Computations were carried
out on XC30 at NAOJ.

\end{document}